\documentclass[prx,twocolumn,floatix,amssymb,amsmath, superscriptaddress,longbibliography]{revtex4-2}

\usepackage{graphicx}
\usepackage{epsfig}
\usepackage{amsmath,amssymb,amsthm}
\usepackage[utf8]{inputenc}
\usepackage[dvipsnames]{xcolor}
\usepackage[colorlinks, urlcolor=blue, linkcolor=red]{hyperref}
\usepackage{soul,xcolor,braket}
\usepackage{comment}
\usepackage{array}
\usepackage{cases}

\setstcolor{red}

\newcommand{\beqa}{\begin{eqnarray}}
\newcommand{\eeqa}{\end{eqnarray}}

\newcommand{\newtext}[1]{\textcolor{MidnightBlue}{ #1}}

\renewcommand{\boxed}[2]{\textcolor{#1}{%
\tikz[baseline={([yshift=-1ex]current bounding box.center)}] \node [rectangle, minimum width=1ex,rounded corners,draw] {\normalcolor\m@th$\displaystyle#2$};}}

\newcounter{appsection}
\newcounter{appsubsection}[appsection]

\usepackage{physics}
\usepackage{bookmark}
\usepackage[normalem]{ulem}

\newcommand\redsout{\bgroup\markoverwith{\textcolor{red}{\rule[0.5ex]{2pt}{2pt}}}\ULon}
\newcommand\bluesout{\bgroup\markoverwith{\textcolor{blue}{\rule[0.5ex]{2pt}{2pt}}}\ULon}
\newcommand\greensout{\bgroup\markoverwith{\textcolor{green}{\rule[0.5ex]{2pt}{2pt}}}\ULon}

\begin{document}

\title{Collective quantum enhancement in critical quantum sensing}
\author{Uesli Alushi}
\email{uesli.alushi@aalto.fi}
\affiliation{Department of Information and Communications Engineering, Aalto University, Espoo, 02150, Finland}
\affiliation{Institute for Complex Systems, National Research Council (ISC-CNR), Via dei Taurini 19, 00185 Rome, Italy}
\author{Alessandro Coppo}
\affiliation{Institute for Complex Systems, National Research Council (ISC-CNR), Via dei Taurini 19, 00185 Rome, Italy}
\affiliation{Physics Department, Sapienza University, P.le A. Moro 2, 00185 Rome, Italy}
\author{Valentina Brosco}
\affiliation{Institute for Complex Systems, National Research Council (ISC-CNR), Via dei Taurini 19, 00185 Rome, Italy}
\affiliation{Physics Department, Sapienza University, P.le A. Moro 2, 00185 Rome, Italy}
\author{Roberto Di Candia}
\email{rob.dicandia@gmail.com}
\affiliation{Department of Information and Communications Engineering, Aalto University, Espoo, 02150, Finland}
\affiliation{Dipartimento di Fisica, Universit\`a degli Studi di Pavia, Via Agostino Bassi 6, I-27100, Pavia, Italy}
\author{Simone Felicetti}
\email{felicetti.simone@gmail.com}
\affiliation{Institute for Complex Systems, National Research Council (ISC-CNR), Via dei Taurini 19, 00185 Rome, Italy}
\affiliation{Physics Department, Sapienza University, P.le A. Moro 2, 00185 Rome, Italy}
\begin{abstract}
Critical systems represent a valuable resource in quantum sensing and metrology. Critical quantum sensing (CQS) protocols can be realized using finite-component phase transitions, where criticality arises from the rescaling of system parameters rather than the thermodynamic limit. Here, we show that a collective quantum advantage can be achieved in a multipartite CQS protocol using a chain of parametrically coupled critical resonators in the weak-nonlinearity limit. We derive analytical solutions for the low-energy spectrum of this unconventional  quantum many-body system, which is composed of locally critical elements. We then assess the scaling of the quantum Fisher information with respect to fundamental resources. We demonstrate that the coupled chain outperforms an equivalent ensemble of independent critical sensors, achieving quadratic scaling in the number of resonators. Finally, we show that even with finite Kerr nonlinearity or Markovian dissipation, the critical chain retains its advantage, making it relevant for implementing quantum sensors with current microwave superconducting technologies.
\end{abstract}

\maketitle

\section{Introduction}
Critical quantum sensing (CQS) is by now an established approach, based on the exploitation of quantum properties spontaneously developed in proximity of phase transitions. In the context of quantum metrology, the performance of parameter estimation protocols~\cite{RevModPhys_QSensing} is assessed based on the precision that can be achieved using a limited amount of fundamental resources, such as the size of the probe system and/or the protocol duration time. By using non-classical properties such as entanglement and superposition, it is possible to overcome the performance of any sensing strategy based on purely classical resources~\cite{Paris2009}. Numerous theoretical studies have shown that a quantum-enhanced sensing precision can be achieved exploiting static~\cite{Zanardi2008,ivanov_adiabatic_2013,Bina2016,Lorenzo2017,Ivanov2020,invernizzi2008Optimal,Mirkhalaf2020,Niezgoda2021,DiFresco2022,DiFresco2024,Sahoo24} or dynamical~\cite{Tsang2013,macieszczak_dynamical_2016,Cabot24,zicari2024} properties of many-body systems in proximity of the critical point. Despite critical quantum systems becoming divergingly slow as they approach the phase transition, the optimal precision scaling can, in principle, be saturated with respect to both probe size and time~\cite{Rams2018}. Furthermore, even if prior information on the parameter to be estimated is required to operate the sensor close to the critical point, CQS protocols work well also in a global-sensing scenario when adaptive strategies are applied~\cite{Montenegro2021,Salvia2023, PhysRevLett.133.120601}.
First implementations with  Rydberg atoms~\cite{Ding2022}, nuclear magnetic resonance~\cite{Liu2021} and superconducting quantum circuits~\cite{Petrovnin_2024,beaulieu2024} demonstrate the experimental feasibility of the CQS approach.\\
\indent Recently, it has been shown~\cite{Garbe2020} that CQS protocols can also be conceived using finite-component phase transitions (FCPTs), where the thermodynamic limit is replaced with a rescaling of the system parameters~\cite{hwang_quantum_2015,Felicetti2020}. This class of phase transitions can emerge in 
quantum resonators with atomic~\cite{Ashhab2013, hwang_quantum_2015,Puebla2017,Peng2019,Zhu2020,Zhang24} or Kerr-like~\cite{Bartolo2016,Felicetti2020,Minganti2023a,Minganti2023b} nonlinearities, and it is of high relevance for CQS for two main reasons: (i) It provides a tractable theoretical framework to analyze the fundamental precision bounds and to design optimal parameter estimation protocols. For instance, FCPTs have been used to demonstrate the intrinsic constant-factor advantage of dynamical over statical CQS protocols~\cite{Chu2021,Garbe2022} and the presence of apparent super-Heisenberg scalings when focusing on a single resource~\cite{Gietka2022,Gietka2022a,Garbe2022a}. Continuous-measurement schemes~\cite{Ilias2022,Yang2022}  have been designed to efficiently retrieve information in the dissipative case, where the optimal precision bounds are in general not achievable. In the presence of a thermal bath, CQS outperforms passive strategies when preparation and measurement times are not negligible~\cite{alushi2024optimality}.
(ii) As demonstrated by recent experiments, FCPTs can be \emph{controllably} implemented using atomic~\cite{cai2021observation},  polaritonic~\cite{DelteilNatMat19,Zejian2022} and circuit-QED ~\cite{Fink2017,Brooks2021,beaulieu2023observation,chen2023quantum,Sett24,jouanny2024} platforms. Accordingly, CQS protocols can be implemented with small-scale devices, without requiring the implementation and control of many-body quantum systems. CQS protocols based on FCPTs have been designed for quantum resonators~\cite{heugel2020_quantum,DiCandia2023,Rinaldi2021,Petrovnin_2024,Hotter24,chen2024critical,choi2024observing}, single trapped-ions~\cite{Ilias2023}, optomechanical~\cite{Bin2019,Tang2023} or magnomechanical~\cite{Wan2023} devices, spin impurities~\cite{mih23multiparameter,mihailescu2024} and Rabi-like systems~\cite{Ying2022,Xie2022,Lu2022,Zhu24}. Finally, physical devices undergoing FCPTs can be interconnected to form one- or two-dimensional arrays of critical systems, thereby extending CQS approaches to complex extended systems.\\
\indent Here, we theoretically show that collective quantum enhancement in CQS can be achieved using an array of coupled critical quantum sensors. In particular, we consider a model composed of a chain of nonlinear quantum resonators coupled via a parametric nearest-neighbor interaction. We focus on the weak-nonlinearity limit, where each resonator of the array can become locally critical and undergo an FCPT. We assess the metrological performance of a static CQS protocol by analytically evaluating the quantum Fisher information (QFI) over the ground state manifold, under Gaussian approximation.  In proximity to the critical point, which is the optimal working regime, we find closed-form analytical solutions for the asymptotic scaling of the QFI with respect to fundamental resources, i.e., the number of resonators, the number of excitations and the protocol-duration time. This analysis allows us to benchmark the estimation precision achievable with the coupled-resonator chain against that of an equal number of independent critical sensors. We find that the coupled chain yields a quadratic enhancement of the QFI scaling with respect to the number of resonators, compared to the linear scaling observed in an uncoupled array. We demonstrate that this conceptual result still holds when including physically relevant corrections to the minimal model.
We first develop a perturbation theory to estimate the saturation point due to local Kerr nonlinearities, and find that the resonator chain presents a practically relevant advantage for finite values of physical parameters. We then show that, using a dynamical CQS protocol, it is still possible to achieve collective quantum advantage in an open-quantum-system setting. We unveil a non-trivial transition between a short-time unitary regime and the stationary regime achieved for long evolution times.
\begin{figure*}[t!]
    \centering
    \includegraphics[width=0.93\textwidth]{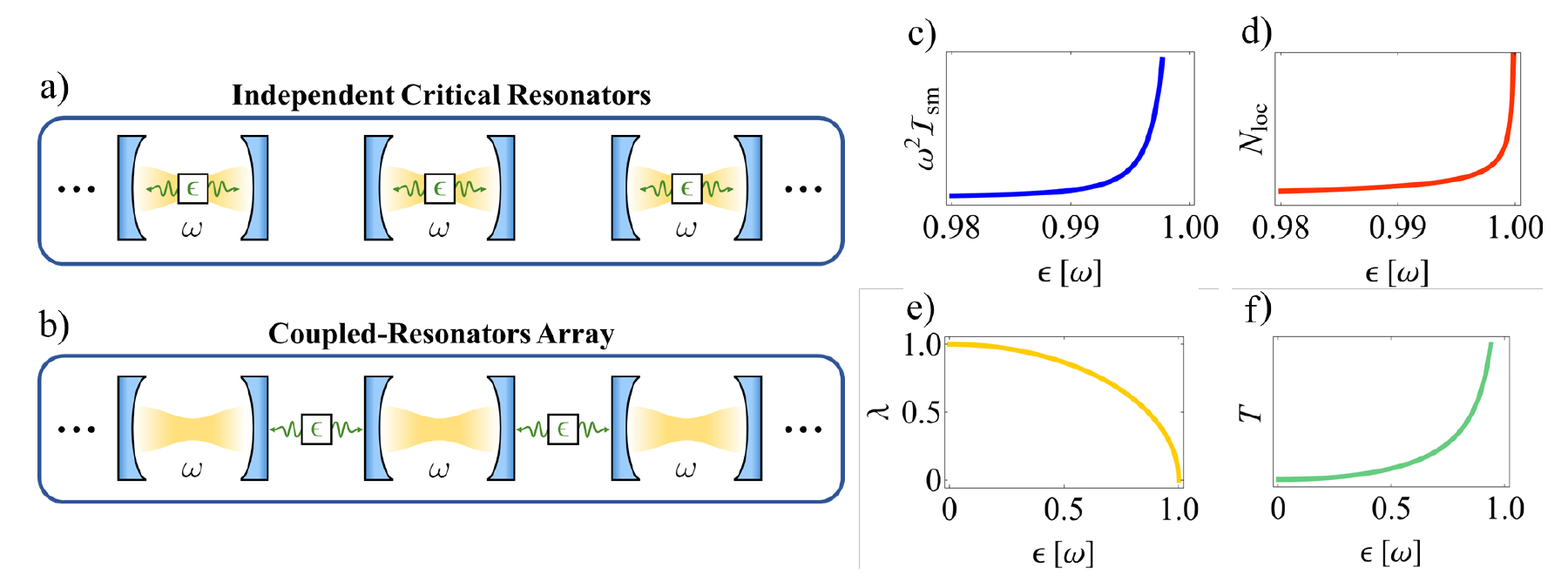}
    \caption{\textbf{Schematic representation of independent critical resonators and the  critical chain.} We illustrate \textbf{a)} an ensemble of independent critical resonators and \textbf{b)} a chain of parametrically-coupled critical resonators.  A single critical sensor is described by the Hamiltonian \eqref{Kerr}, while the coupled-resonator chain is modeled by the Hamiltonian \eqref{Hm_complete}. Focusing on a single independent critical sensor, we plot \textbf{c)} the QFI \(\mathcal{I_{\rm sm}}\) and \textbf{d)} the local number of photons \(N_{\rm loc}\) as functions of \(\epsilon\). As the squeezing parameter \(\epsilon\) approaches the critical point \(\epsilon_{\rm c}=\omega\), both \(\mathcal{I}_{\rm sm}\) and \(N_{\rm loc}\) diverge. Furthermore, we show \textbf{e)} the energy gap \(\lambda\) and \textbf{f)} the estimate of the protocol duration time \(T\) as a function of \(\epsilon\). When \(\epsilon\to\epsilon_{\rm c}\), \(\lambda\) tends to zero, indicating that the energy gap between the ground and first-excited state closes. Consequently, the time required to perform the adiabatic sweep diverges. To obtain the plot for \(T\) we set \(\eta=1\).  }
    \label{Sketch}
\end{figure*}
\section{SUMMARY OF RESULTS}\label{SII}
\subsection{Critical quantum sensing: adiabatic protocol}  
Here, we briefly introduce the CQS protocol and the methods used to assess parameter-estimation performance.
We consider as a sensor a system whose Hamiltonian depends on an unknown parameter $\omega$ to be estimated, while all other parameters are known, and such that one control parameter $\epsilon$ can be tuned in real-time. A static CQS protocol can be defined as a three-step algorithm: (i) the system is initialized in its ground state in a trivial region of the parameter space, say $\epsilon=0$; (ii) the control parameter $\epsilon$ is adiabatically tuned close to the critical value $\epsilon_{\rm c}$, slowly enough so that the system never leaves its instantaneous ground state; (iii) a measurement is performed on the final state, and the outcome is used to infer information about the value of $\omega$. The working principle behind adiabatic CQS protocols is the exploitation of the high susceptibility of the system ground state with respect to small variations of the parameter to be estimated. 

 The ultimate precision achievable in parameter estimation is given by the quantum Cram{é}r-Rao bound~\cite{Paris2009}. The mean squared error is lower-bounded by \(\Delta\omega^2\geq\left(M\mathcal{I}\right)^{-1}\), where \(M\) is the number of independent measurements and \(\mathcal{I}\) is the QFI. For our adiabatic protocol, the QFI must be evaluated [as in Eq.~\eqref{qfi}] on the manifold defined by the system ground state as the parameter $\omega$ is varied. The QFI represents the maximum amount of information that can be extracted on average for a given repetition of the protocol, assuming that the optimal observable is measured after the adiabatic sweep. The performance of a parameter-estimation protocol can be assessed by analyzing the scaling of the QFI with the fundamental resources used during the protocol, e.g., the size of the probe system \(N\). For example, in Ramsey interferometry, the QFI for phase estimation can grow at most linearly as \(\mathcal{I}\sim N\) when the spins are in a separable state, while it can achieve the quadratic scaling \(\mathcal{I}\sim N^2\) when the spins are entangled~\cite{Maccone_entangl}. This quadratic enhancement is dubbed Heisenberg scaling and is achievable only using nonclassical resources such as entanglement or squeezing.

The size of the probe system is not the only relevant resource. When dealing with critical systems the protocol time must unavoidably be taken into account as a fundamental resource. When the critical point is approached, the energy gap between the ground and first-excited state tends to zero, and so the time \(T\) required to perform an adiabatic sweep unavoidably diverges. Analyzing the scaling of the QFI only with respect to the number of probe systems can then lead to an apparent super-Heisenberg scaling. Once time is taken into account~\cite{Rams2018,Garbe2020}, under very general assumptions, the optimal Heisenberg scaling is given by $\mathcal{I}\sim T^2 N^2$. However, this is strictly valid only for closed quantum systems. When decoherence and dissipation are included, stricter asymptotic bounds apply~\cite{Gorecki_bounds,demkowicz2017adaptive, Gorecki24}. Here, we limit ourselves to the unitary case, as our focus is on the collective enhancement achievable with quantum-correlated probes, and our results can be generalized to the dissipative case.
\subsection{Critical parametric quantum sensor} 
Let us now introduce our physical model of a critical quantum sensor, which is a quantum nonlinear resonator with two-photon pumping terms, described by the Hamiltonian
\begin{equation}\label{Kerr}
    H=\omega a_1^\dag a_1+\frac{\epsilon}{2}\left(a_1^2+a_1^{\dag2}\right)\, + \chi a_1^\dag a_1^\dag a_1 a_1\,.
\end{equation}
Here and throughout the manuscript, we set \(\hbar=1\).
This system undergoes an FCPT in the limit $\chi \to 0$, when \(\epsilon\) approaches the critical point $\epsilon_{\rm c} = \omega$. This model is of direct experimental relevance~\cite{beaulieu2023observation}, and it effectively describes the critical scaling of fully-connected models~\cite{Garbe2022}, a broad class of quantum-optical critical systems. The CQS protocol consists of initializing the system in the vacuum state, and adiabatically tuning $\epsilon$ to a value close to $\epsilon_{\rm c}$. To work in this critical regime, we assume to have good prior information on the parameter to be estimated, namely $\omega = \omega_0 + \delta \omega$, where $\omega_0$ is known and $\delta \omega\ll \omega_0$ is an unknown frequency shift to be estimated. For global quantum sensing, i.e., when prior information is not available, efficient adaptive strategies can be implemented~\cite{Montenegro2021,Salvia2023}. 

Under Gaussian approximation (\(\chi=0\)),
we can obtain analytical results. Indeed, we first compute the ground state manifold, and then evaluate the QFI for the estimation of $\omega$. Consequently, we provide an upper-bound on the estimation precision with respect to the involved resources, i.e., the number of photons \(N_{\rm loc}=\langle a_1^\dagger a_1\rangle\) and the protocol duration time \(T\). Here, the subscript stands for \emph{local} and will be meaningful in the multipartite case.
By applying time-dependent perturbation theory, it can be shown~\cite{Garbe2020} that the time required to perform the adiabatic sweep is of the order of $T\sim (\eta \lambda)^{-1}$, where \(\lambda\) is the energy gap between the ground and first-excited state, and $\eta<1$ is a parameter which controls the adiabaticity of the protocol. 
As shown in Fig.~\ref{Sketch}, when approaching the critical point the QFI and the photon number diverge, while the energy gap between the ground and the first-excited state vanishes, as expected. For a single-mode critical sensor, the QFI scales as $\mathcal{I}_{\rm sm}\sim 2\eta^2 T_{}^2N_{\rm loc}^2$. Here, the subscript ``sm" stands for ``single-mode". We stress that in the limit \(\chi\to0\), this result can be analytically derived. Notice that in any physical implementation nonlinear terms are always present, and theoretically they are required to obtain a stable dynamics in the whole parameter space~\cite{Bartolo2016}. In the case of finite nonlinearity, this scaling is valid until the photon number becomes large enough to break the Gaussian approximation and reach saturation. As shown in Appendix \ref{APPC}, perturbation theory reveals that the saturation point occurs when $N_{\rm loc}\sim \sqrt[3]{\omega/132\chi}$.
\subsection{Collective quantum advantage in CQS} 
In this work, we perform a comparison between an ensemble of $M$ independent critical sensors, with a chain of parametrically-coupled resonators (see Fig.~\ref{Sketch}). The QFI is additive when using \(M\) independent sensors or, equivalently, \(M\) uncorrelated repetitions of the parameter-estimation protocol. Accordingly, when $M$ independent critical sensors are used, we straightforwardly obtain $\mathcal{I}_{\rm ind}\sim 2 M \eta^2 T_{}^2N_{\rm loc}^2$, with the subscript ``ind" standing for ``independent". This result can also be rewritten in terms of the total number of photons $N = M N_{\rm loc}$, as $\mathcal{I}_{\rm ind}\sim 2 \eta^2 T_{}^2 N_{}^2/M$.\\ 
\indent Our collective critical sensor is defined as a coupled-resonator chain with Hamiltonian
\begin{equation}\label{Hm_complete}
H= \sum^{M}_{j=1} \left[
\omega a^\dag_ja_j + \frac{\epsilon}{2} \left( a_ja_{j+1}+a^\dag_ja^\dag_{j+1} \right)
+ \chi a^\dag_ja^\dag_ja_j a_j
\right].
\end{equation} 

\noindent We derive analytical solutions for the QFI for the estimation of $\omega$, as well as for the photon number and the energy gap, under Gaussian approximation. The result for the QFI is shown in Eqs.~\eqref{QFIE} and \eqref{QFIO} for even and odd \(M\), respectively. We then obtain approximate closed-form analytical solutions in two relevant limits: in proximity of the phase transition and in the continuous limit (large $M$). The analysis unveils interesting features of this many-body model composed of locally critical constituents. In this summary, we focus on the most effective regime for quantum sensing, which is the critical scaling of a chain with an odd number of resonators. The QFI achievable in this case scales as $I_O = 2\eta^2 M^2 T^2 N_{\rm loc}^2 = 2\eta^2 T^2 N^2$. Thus, the critical chain presents a quadratic enhancement in the number of resonators $M$ with respect to the non-interacting case. The critical chain is advantageous even when considering a finite value of the Kerr nonlinearity. Indeed, it can be effectively described by a single-mode resonator with a diluted nonlinearity $\chi/M$ (see Appendix \ref{APPD}), and the saturation point is reached when $N\sim\sqrt[3]{M \omega/132 \chi}$. Finally, the collective quantum advantage can also be achieved in a dissipative setting, when a dynamical protocol is used. Different regimes can be identified with respect to the timescale set by $3/(4M N_{\rm loc} \Gamma)$, where $\Gamma$ is the dissipation rate. For a very short evolution time, the dynamics is effectively unitary, and a scaling advantage is still achievable. Conversely, for a very long evolution time the collective quantum advantage fades away, in accordance with fundamental bounds~\cite{Gorecki_bounds, demkowicz2017adaptive,Gorecki24}. We also find a non-trivial transient regime characterized by a collective constant-factor advantage.

\section{Local models}\label{SIII}
In this section, we analyze the performance of single-and two-mode critical quantum sensors. In both cases, we compute the QFI and express it in terms of the resources involved, i.e., the average number of photons in each mode and the duration time of the sensing protocol. The results will be used as a benchmark, as well as an intermediate step to solve the interacting chain.
\subsection{Single-mode critical quantum system}\label{SIIIA}
A single-mode critical quantum sensor can be modeled, under Gaussian approximation, by the Hamiltonian
\begin{equation}\label{Hsm}
    H_{\rm sm}=\omega a_1^\dag a_1+\frac{\epsilon}{2}\left(a^2_1+a^{\dag2}_1\right)\,,
\end{equation}
where \(\epsilon\) is the squeezing parameter, \(\omega\) is the mode frequency and \(a_1\) is the bosonic mode. The full model presented in Eq.~\eqref{Kerr} undergoes a second-order phase transition in proximity of the critical point \(\epsilon_{\rm c}=\omega\) in the limit $\chi \to 0$. For \(\epsilon<\epsilon_{\rm c}\) the model is well approximated by the Hamiltonian \eqref{Hsm}, which is diagonalizable via a Bogoliubov Transformation (BT), obtaining 
\begin{equation}\label{hsm}
    H_{\rm sm}=\lambda d_1^\dag d_1+\frac{1}{2}\left(\lambda-\omega\right)\,.
\end{equation}
We defined \(\lambda=\sqrt{\omega^2-\epsilon^2}\), and \(d_1\) is the normal mode related to \(a_1\) via the transformation
\begin{equation}\label{newmodes1}
    a_1=td_1-sd_1^\dag\,,
\end{equation}
where the parameters \(t\) and \(s\) are
\begin{align}
 t&=\frac{(\lambda+\omega)}{\sqrt{2(\omega^2-\epsilon^2)+2\omega\lambda}}\,,\\\quad
    s&=\frac{\epsilon}{\sqrt{2(\omega^2-\epsilon^2)+2\omega\lambda}}\,.\label{s}   
\end{align}
 Let us focus now on the estimation of the parameter \(\omega\). Given a manifold of pure states $\ket{\psi}$, the QFI can be evaluated as~\cite{Paris2009}
\begin{equation}\label{qfi}
\mathcal{I}=4\left(\bra{\partial_\omega\psi}\ket{\partial_\omega\psi}-|\bra{\partial_\omega\psi}\ket{\psi}|^2\right)\,,
\end{equation}
where  \(\ket{\partial_\omega\psi}\) is the partial derivative with respect to the parameter to be estimated. To assess the performance of the adiabatic CQS protocol, we will compute the QFI associated with the system ground state \(\ket{g}\). The ground state of the Hamiltonian \eqref{Hsm} is the single-mode squeezed vacuum \(\ket{g}=S_1(\xi)\ket{0}\). We define the  single-mode squeezing operator \(S_i(\xi)=e^{\frac{1}{2}\left(\xi^*a_i^2-\xi a_i^{\dag 2}\right)}\) and the squeezing parameter \(\xi=|\xi|=\ln{\left(s+\sqrt{s^2+1}\right)}\).
This state has an average number of photons \(N_{\rm loc}=\sinh^2(|\xi|)=s^2\), where ``loc" stands for ``local".
By inserting in \eqref{qfi} the expression of the system ground state, we obtain the single-mode QFI for the estimation of \(\omega\):
\begin{equation}\label{Ism}
\mathcal{I}_{\rm sm}=\frac{\epsilon^2}{2(\omega^2-\epsilon^2)^2}\,.
\end{equation}
The dominant contribution to the duration of the sensing protocol is given by the time $T$ required to perform the adiabatic sweep, which must be larger~\cite{Garbe2020} than the inverse of the gap between the ground and first-excited state. From \eqref{hsm}, it immediately follows that the energy gap is \(\lambda=\sqrt{\omega^2-\epsilon^2}\). Consequently, the protocol duration time can be expressed as \(T\approx\ (\eta \lambda)^{-1}\), with $\eta <1$. To analyze the critical scaling approached when \(\epsilon\to\epsilon_{\rm c}\), we define \(\epsilon=(1-x)\omega\) with \(0<x<1\), and compute the QFI in the asymptotic regime \(x\to0\). Close to the critical point, we have \(\mathcal{I}_{\rm sm}\approx(2\sqrt{2}\omega x)^{-2}\), \(T\approx(\omega\sqrt{2x})^{-1}\) and \(N_{\rm loc}\approx(8x)^{-\frac{1}{2}}\). Using the latter, we express \(x\) in terms of the number of photons as \(x\approx(2\sqrt{2}N_{\rm loc})^{-2}\), which implies \(T_{}\approx2N_{\rm loc}/\eta \omega\). Finally, in terms of the resources involved the single-mode QFI scales as
\begin{equation}\label{ISMscaling}
\mathcal{I}_{\rm sm}\sim2 \eta^2 T_{}^2N_{\rm loc}^2\,.
\end{equation}
\subsection{Two-mode critical quantum system}\label{SIIIB}
Similarly to the single-mode case, a two-mode critical quantum system can be modeled with a two-mode squeezing Hamiltonian.
The single-resonator critical sensor can be generalized to a two-mode critical system as follows:
\begin{equation}
\label{Ham_tm}
    H_{\rm tm}=\omega\left(a_1^\dag a_1+a_2^\dag a_2\right)+\epsilon\left(a_1a_2+a_1^\dag a_2^\dag\right)\,,
\end{equation}
where \(\epsilon\) is the  parametric-coupling strength, \(\omega\) is the frequency of each mode and \(a_1\) and \(a_2\) are two independent bosonic modes. Here, the subscript ``tm" means ``two-mode". Also in this case, the model undergoes a second-order phase transition, when a vanishingly small Kerr nonlinearity is included~\cite{Felicetti2020}. The normalization of the coupling parameter is chosen to have the same critical point \(\epsilon_{\rm c}=\omega\) of the Hamiltonian~\eqref{Kerr}. When \(\epsilon<\epsilon_{\rm c}\), the Hamiltonian \eqref{Ham_tm} can be diagonalized with the following BT:
\begin{align}\label{newmodes2}
    a_1&=td_1-sd_2^\dag\,,\nonumber\\\quad
    a_2&=td_2-sd_1^\dag\,,
\end{align}
where \(d_1,\,d_2\) are the normal modes. We can then write
\begin{equation}
    H_{\rm tm}=\lambda(d_1^\dag d_1+d_2^\dag d_2)+\lambda-\omega\,.
\end{equation}
The ground state of the system is a two-mode squeezed vacuum state \(\ket{g}=S_{1,2}(\xi)\ket{0}\), with \(S_{i,j}(\xi)=e^{\left(\xi^*a_ia_j-\xi a_i^{\dag }a_j^\dag\right)}\) a two-mode squeezing operator. Notice that the squeezing parameter \(\xi=|\xi|\) is the same as in the single-mode case.  The average number of photons in the ground state is \(N_{}=2\sinh^2(|\xi|)=2s^2\) since now there are two normal modes with the same eigenvalue $\lambda$ and the same number of photons \(N_{\rm loc}= \langle a_1^\dagger a_1\rangle= \langle a_2^\dagger a_2\rangle= s^2\). Using \eqref{qfi}, we can compute the two-mode QFI for the estimation of \(\omega\):
\begin{equation}\label{Itm}
\mathcal{I}_{\rm tm}=\frac{\epsilon^2}{(\omega^2-\epsilon^2)^2}\,.
\end{equation}
In the asymptotic limit \(x\to0\), we have \(\mathcal{I}_{\rm tm}\approx(2\omega x)^{-2}\) and \(N_{\rm loc}\approx(8x)^{-\frac{1}{2}}\). Since the energy gap is still \(\lambda\), the protocol duration time remains unchanged with respect to the single mode case. Indeed, expressing \(x\) as \(x\approx(2\sqrt{2}N_{\rm loc})^{-2}\) leads to \(T_{}\approx2N_{\rm loc}/\eta\omega\). Finally, in terms of the involved resources the two-mode QFI scales as
\begin{equation}
\mathcal{I}_{\rm tm}\sim4 \eta^2 T_{}^2N_{\rm loc}^2\,.
\end{equation}
Comparing this result with Eq.~\eqref{ISMscaling}, we see that $\mathcal{I}_{\rm tm}$  is simply twice what is obtained with a single critical sensor. A two-resonator chain is 
 then equivalent to two independent resonators, 
suggesting that there is no advantage in considering coupled critical sensors. However, in the following, we demonstrate that this conclusion does not hold for larger chains.
\section{Critical resonator chain}\label{SIV}
Let us finally consider the critical resonator chain. Under Gaussian approximation, it is described by the Hamiltonian
\begin{equation}\label{Hm}
H= \sum^{M}_{j=1} \left[
\omega a^\dag_ja_j + \frac{\epsilon}{2} \left( a_ja_{j+1}+a^\dag_ja^\dag_{j+1} \right)\right]\,.
\end{equation} 
 Assuming periodic boundary conditions \(a_{M+j}=a_j\), the Hamiltonian \eqref{Hm} is diagonalizable in the reciprocal space. By applying the discrete Fourier transform  \(a_j=\left(M\right)^{-\frac{1}{2}}\sum_ka_ke^{-ikj}\), we obtain 
\begin{align}\label{Hmk}
H&=H_0+\delta_{M,2m}H_{\pi}+\sum_{k>0}\omega\left(a^\dag_k a_k+a^\dag_{-k} a_{-k}\right)+\nonumber\\\quad&+\sum_{k>0}\epsilon\cos{(k)}\left(a_{k}a_{-k}+a^\dag_{k}a^\dag_{-k}\right)\,.
\end{align}
Here, \(\delta_{M,2m}\) is a Kronecker delta which is nonzero only for even $M$. The terms \(H_0,\,H_\pi\) are single-resonator critical models defined as in Eq.~\eqref{Hsm}, but replacing $a_1$ with \(a_0\) and \(a_{\pi}\), respectively. The sign of the coupling for the \(a_{\pi}\) mode must be inverted, a phase difference that has no consequence for the current analysis. The periodic boundary conditions imply that we can restrict ourselves to the first Brillouin zone (FBZ), and
\(k=2n\pi/M\), with \(n\in\mathbb{Z}\). The FBZ is defined as follows: if \(M\) is even, \(n\in[-M/2+1,\,M/2]\); if \(M\) is odd, \(n\in[-(M-1)/2,\,(M-1)/2]\). We point out that the mode \(a_{\pi}\) is present only in the case in which the number of modes \(M\) is even. \\
\indent In reciprocal space, the multi-mode Hamiltonian \eqref{Hmk} consists of \(M\) independent critical systems. Specifically, two single-mode systems defined by \(a_0\) and \(a_{\pi}\), along with a series of two-mode systems defined by pairs \(\{a_k,\,a_{-k}\}\). Each element can be independently diagonalized using the single- and two-mode BTs derived previously, by replacing \(\epsilon\to\epsilon\cos{(k)}\). Notice that the BT can be performed only if \(\epsilon<\epsilon_{\rm c}\). In the multi-mode case, we have many critical points defined by the dispersion relation \(\epsilon_{\rm c}=\omega/\cos(k)\). This means that the multi-mode Hamiltonian is diagonalizable via a BT only if the parameter \(\epsilon\) is smaller than the lowest among the critical points. Therefore, we restrict ourselves to the case \(\epsilon<\omega\). If this condition is satisfied, applying the BT to \eqref{Hmk} results in
\begin{align}\label{HKdiagonal}
    H&=\sum_{k\in FBZ}\lambda_k\left(d^\dag_k d_k+\frac{1}{2}\right)-\frac{M}{2}\omega\,,
\end{align}
where \(\lambda_k=\sqrt{\omega^2-\epsilon^2\cos^2{(k)}}\), and \(d_k\) are independent normal modes. The modes \(d_0,\,d_\pi\) are defined as in \eqref{newmodes1}, while the modes \(d_{\pm k}\) as in \eqref{newmodes2}, with the substitution \(\epsilon\to\epsilon\cos(k)\). The ground state of the system is then a multi-mode squeezed vacuum state
\begin{equation}\label{gk}
\ket{g}=S_0(\xi_0)S_{\pi}(\xi_{\pi}) \bigotimes_{k>0}S_{k,-k}(\xi_{k})\ket{0}\,.
\end{equation}
 We recall that \(S_i(\xi_i)=e^{\frac{1}{2}\left(\xi_i^*a_i^2-\xi_i a_i^{\dag 2}\right)}\) with \(i=\{0,\pi\}\) is a single-mode squeezing operator associated with the modes \(a_0,\,a_{\pi}\). Conversely, \(S_{k,-k}(\xi_k)=e^{\left(\xi_k^*a_ka_{-k}-\xi_k a_k^{\dag }a_{-k}^\dag\right)}\) is a two-mode squeezing operator related to pairs of coupled modes \(\{a_k,\,a_{-k}\}\). The squeezing parameters are \(\xi_k=|\xi_k|=\ln{\left(s_k+\sqrt{s_k^2+1}\right)}\) for \(k\in[0,\pi/2)\) and \(\xi_{k}=|\xi_k|e^{i\pi}\) for \(k\in[\pi/2,\pi]\). The parameter \(s_k\) is obtained from \eqref{s} with the substitution \(\epsilon\to\epsilon\cos(k)\). The total average number of photons in the multi-mode ground state is given simply by the sum of the photons in each mode \(a_k\). Given that the ground state is a multi-mode squeezed vacuum state, each mode contains \(N_k=\sinh^2(|\xi_k|)=s_k^2\) photons. Since \(s^2_k=s^2_{-k}\), the total number of photons in the ground state can be expressed as  
\begin{equation}\label{photonMM}
N=\sum_{k}N_k=s_0^2+\delta_{M,2m}s_{\pi}^2+\sum_{k>0}2s_k^2\,.
\end{equation}
Moreover, the local number of photons at each site \(j\) is determined by \(N_{\rm loc}=\langle a^\dag_j a_j\rangle= \sum_k \langle a_k^\dag a_k\rangle/M=N/M\). A detailed derivation is provided in Appendix \ref{AppendixPhotons} where we show that \(N_{\rm loc}\) is independent of the specific site, i.e., each resonator has the same number of photons.
\begin{figure}[t!]
  \includegraphics[width=0.48
\textwidth]{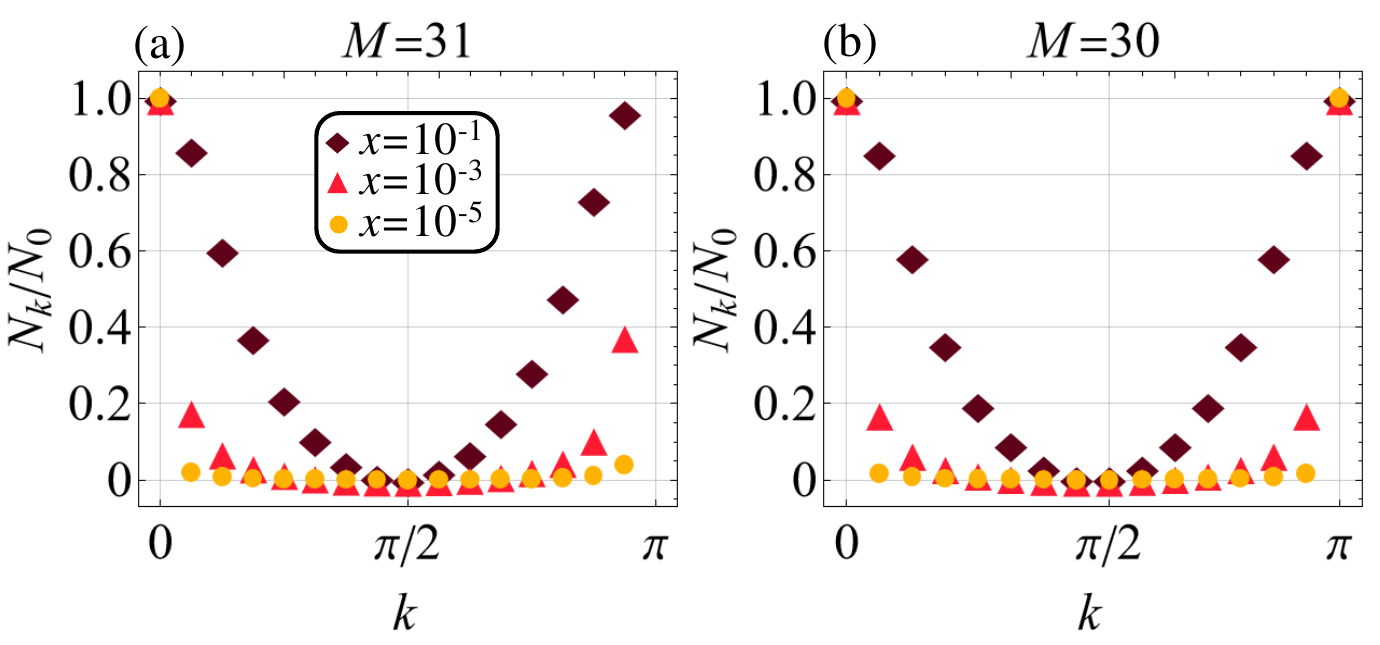}
\centering
\caption{\textbf{Local number of photons.} We plot the ratio \(N_k/N_0\) for a number of resonators \textbf{a)} \(M=31\) and \textbf{b)} \(M=30\) for different values of \(x=1-\epsilon/\omega\). As the critical point is approached ($x\to 0$), the majority of photons is strongly confined in the mode \(a_0\) when \(M\) is odd, and in the modes \(a_0,\,a_{\pi}\) when \(M\) is even (see yellow circles). The fraction of the photons populating the remaining modes is vanishingly small.}
\label{PhotonNumber}
\end{figure}
It is important to stress that in the reciprocal space, when the critical point is approached, the photons are strongly confined only in the mode \(a_0\) when \(M\) is odd and in the modes \(a_0,\,a_{\pi}\) when \(M\) is even. The fraction of photons populating all the other modes vanishes as $\epsilon\to \epsilon_{\rm c}$ (see Fig.~\ref{PhotonNumber}). In this scenario, we can effectively approximate the critical chain as a single-mode system with \(N=MN_{\rm loc}\) photons if \(M\) is odd, or as a two-mode system with \(N=MN_{\rm loc}/2\) photons per mode when \(M\) is even.\\ 
Having obtained its ground state, we can now use \eqref{qfi} to calculate the QFI of the critical chain. From Eq.~\eqref{gk} we see that the ground state is separable in reciprocal space, hence the QFI of the critical chain is given by the sum of the QFIs of each individual mode $a_0$ and $a_\pi$, and of each pair $\{a_k,a_{-k}\}$. We obtain
\begin{align}
\mathcal{I}_{\rm E}&=\sum_{n=0}^{\frac{M}{2}-1}\frac{\epsilon^2\cos^2{(\frac{2\pi n}{M})}}{\left[\omega^2-\epsilon^2\cos^2{(\frac{2\pi n}{M})}\right]^2}\,,\label{QFIE}\\\quad
\mathcal{I}_{\rm O}&=\frac{\epsilon^2}{2(\omega^2-\epsilon^2)^2}+\sum_{n=1}^{\frac{M-1}{2}}\frac{\epsilon^2\cos^2{(\frac{2\pi n}{M})}}{\left[\omega^2-\epsilon^2\cos^2{(\frac{2\pi n}{M})}\right]^2}\label{QFIO}\,,
\end{align}
where \(\mathcal{I}_{\rm E}\) is the QFI for an even number of modes, while \(\mathcal{I}_{\rm O}\) is the QFI for an odd number of modes. Let us now provide closed-form analytical expressions of the multi-mode QFI in relevant asymptotic regimes. First, we consider a finite number of modes and compute explicitly the QFI close to the critical point. Then, we compute the QFI in the continuous limit $M\to \infty$. In both cases, we express the QFI in terms of the involved resources, i.e., the local number of photons \(N_{\rm loc}\) and the total protocol duration time \(T\). 
\subsection{Asymptotic regimes: Critical scaling}\label{SIVA}
In this section, we provide the scaling behavior of the critical chain QFI in the asymptotic regime where \(\epsilon\to\epsilon_{\rm c}\), assuming a finite \(M\). To analyze the critical scaling we define \(\epsilon=(1-x)\omega\) with \(0<x<1\), where the critical point is approached for \(x\to0\). We then expand Eqs.~\eqref{QFIE} and \eqref{QFIO} in power series, up to the first order in \(x\). Following this expansion, the QFI separates into two terms: the first depends solely on \(x\) and diverges as the critical point is approached, while the second term is a discrete sum dependent only on the number of modes \(M\). This sum can be analytically computed using established results from trigonometric series, allowing us to derive a closed expression for the asymptotic QFI. See Appendix \ref{AppendixSeries} for further mathematical details. Thus, in the limit \(x\to0\), up to the terms of order $x$, the QFI for even and odd critical chains can be expressed  as 
\begin{align}
    \mathcal{I}_{\rm E}&\approx \frac{1}{4\omega^2}\left(\frac{1}{x^2}-\frac{1}{x}-\frac{1}{4}+\frac{M^4-20M^2+64}{180}\right)\,,\label{QFIEasy}\\
    \mathcal{I}_{\rm O}&\approx\frac{1}{8\omega^2}\left(\frac{1}{x^2}-\frac{1}{x}-\frac{1}{4}+\frac{4M^4-20M^2+16}{45}\right)\,.\label{QFIOasy}
\end{align}

\begin{figure*}[ht!]
    \centering
    \includegraphics[width=0.99\textwidth]{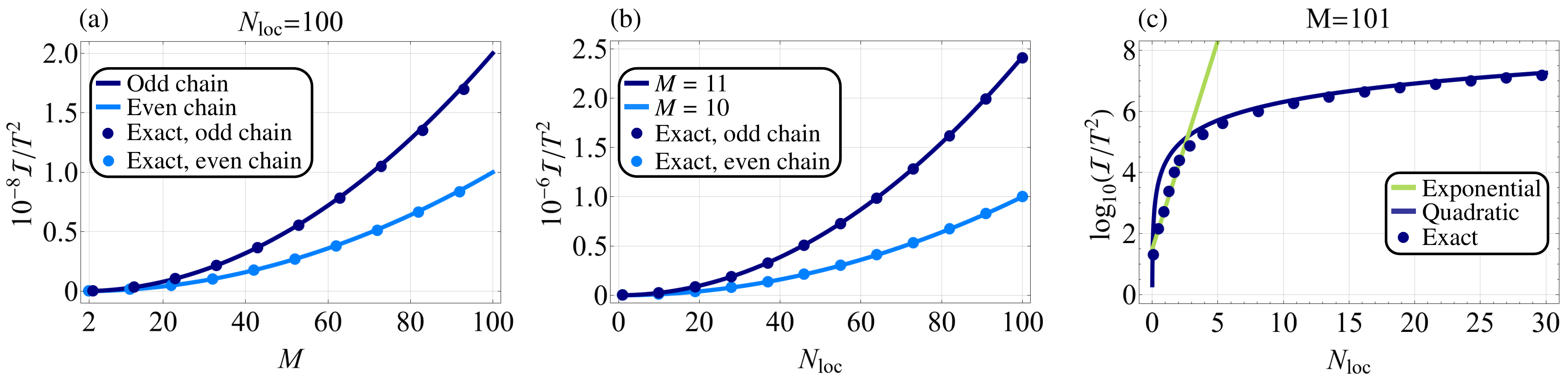}
    \caption{\textbf{Quantum Fisher information}. We show the ratio \(\mathcal{I}/T^2\) as a function of \textbf{a)} the number of modes \(M\), for \(M\) odd (dark blue) and \(M\) even (light blue); and as a function of \textbf{b)} the local number of photons \(N_{\rm loc}\), for \(M=11\) (dark blue) and \(M=10\) (light blue). The quadratic scaling \eqref{QFIEscaling}-\eqref{QFIOscaling} (solid lines) fits well with the exact ratio (small circles) computed numerically from \eqref{QFIE}-\eqref{QFIO}. \textbf{c)} Plot of \(\log_{10}\left(\mathcal{I}/T^2\right)\) as a function of \(N_{\rm loc}\), for \(M=101\). The exact ratio computed numerically (small circles) is well approximated by the exponential scaling \eqref{QFIinfiniteM} (solid green line) only when \(N_{\rm loc}=\mathcal{O}(1)\). However, as soon as \(N_{\rm loc}\) becomes comparable to \(M\), the quadratic scaling \eqref{QFIOscaling} fits well the exact ratio (solid blue line), while the exponential scaling is no longer valid. The same behavior holds also in the case of an even \(M\). To obtain the plots, for each value of \(M\) we have optimized the QFI under the constraint \(N_{\rm loc}=N_{\rm max}\), where in \textbf{a)} \(N_{\rm max}=100\), while in \textbf{b)} \(N_{\rm max}\in(0,100]\) and in \textbf{c)} \(N_{\rm max}\in(0,30]\). We have set \(\omega=1\).
    }
    \label{QFIfigures}
\end{figure*}

A similar analysis applies to the average number of photons in the multi-mode ground state.  Notice that the expression for the number of photons is given by \eqref{photonMM}, and the only distinction between odd and even $M$ lies in the range of \(k\) values in the discrete sum. Then, expanding \eqref{photonMM} up to terms of order $\sqrt{x}$ yields
\begin{numcases}{N\approx}
\frac{1}{\sqrt{2x}}+\frac{M}{\pi}\left[\log\left(\frac{M}{\pi}\right)+\gamma-\frac{\pi}{2}\right], 
\;\, M\ \text{even} \label{NE}
\\
\frac{1}{\sqrt{8x}}+\frac{M}{\pi}\left[\log\left(\frac{2M}{\pi}\right)+\gamma-\frac{\pi}{2}\right], 
\;\, M\ \text{odd} \label{NO}
\end{numcases}
where \(\gamma\) is the Euler-Mascheroni constant. See Appendix \ref{AppendixSeries} for further details on the derivation of Eqs.~\eqref{NE}-\eqref{NO}. Our aim now is to express the multi-mode QFI in terms of the resources involved, namely the local number of photons \(N_{\rm loc}\) and the protocol duration time \(T\). To achieve this, we first express the parameter \(x\) as a function of the local number of photons using \eqref{NE}-\eqref{NO}. Specifically, we find \(x\approx(\sqrt{2} M N_{\rm loc})^{-2}\) when \(M\) is even and \(x\approx(2\sqrt{2} M N_{\rm loc})^{-2}\) when \(M\) is odd. It is important to note that these relations hold true under the condition that the local number of photons is sufficiently large, satisfying \(\pi N_{\rm loc}\gg \left|\ln{\left(M/\pi\right)+\gamma-\pi/2}\right|\) for $M$ even, and 
\(\pi N_{\rm loc}\gg \left|\ln{\left(2M/\pi\right)+\gamma-\pi/2}\right| \) for $M$ odd.
This assumption is reasonable in the considered asymptotic regime, where \(M\) is finite and near the criticality. On the other hand, from \eqref{HKdiagonal} we have that the lowest energy gap is \(\lambda_0\), which means that the protocol duration time is estimated as \(T\approx(\eta\lambda_0)^{-1}\approx(\eta\omega\sqrt{2x})^{-1}\). In terms of the local number of photons we write \(T\approx MN_{\rm loc}/\eta\omega\) for $M$ even and \(T\approx 2MN_{\rm loc}/\eta\omega\) for $M$ odd. Finally, inserting the expression of \(x\) in \eqref{QFIEasy}-\eqref{QFIOasy} and taking into account the protocol duration time, the QFI scales as
\begin{align}
    \mathcal{I}_{\rm E}&\sim \textcolor{white}{2}\eta^2 M^2T^2N_{\rm loc}^2\,,\label{QFIEscaling}\\\quad
    \mathcal{I}_{\rm O}&\sim 2 \eta^2 M^2T^2N_{\rm loc}^2\,.\label{QFIOscaling}
\end{align}
These solutions then predict a quadratic scaling of the QFI with the number of resonators $M$, which implies a collective quantum enhancement of the estimation precision. The difference between the even and odd cases arises from the fact that, for an odd $M$, only the mode with $k=0$ becomes critical whereas, for an even $M$, both the modes with $k=0$ and $k=\pi$ become critical (see Fig.~\ref{PhotonNumber}). In Fig.~\ref{QFIfigures}(a) and (b), we plot the optimized ratio \(\mathcal{I}/T^2\) as a function of \(M\) and as a function of \(N_{\rm loc}\), respectively. We show that the asymptotic expressions \eqref{QFIEscaling}-\eqref{QFIOscaling} align perfectly with the full analytical solutions of Eqs.~\eqref{QFIE}-\eqref{QFIO}, in the asymptotic regime where \(x\to0\) and \(M\) is finite.
\subsection{Asymptotic regimes: Continuous limit}\label{SIVB} 
As \(M\) increases, the separation between adjacent modes decreases. This is due to the periodic boundary conditions, which impose \(\Delta k=2\pi/M\), so that \(\Delta k\to0\) when \(M\to\infty\). Consequently, in this regime we have a continuum of modes \(a_k\). Therefore, we can compute the QFI in \eqref{QFIE}-\eqref{QFIO} by approximating the discrete sums with definite integrals over this continuum, obtaining
\begin{equation}\label{qfilargeM}
    \mathcal{I}=\frac{M}{4\omega^2}\frac{(1-x)^2}{(x(2-x))^\frac{3}{2}}\,.
\end{equation}
Notice that, of course, by making this approximation, a numerical error is introduced. This error becomes negligible when the number of modes is sufficiently large, such that \(\Delta k\to dk\). We estimate the absolute error incurred when approximating a discrete sum with an integral via
\begin{equation}\label{errore}
    \left|\sum_{k>0}f(k)\Delta k-\int_0^{\pi}f(k)dk\right|< \frac{M}{2}\Delta k^2\text{max}\{\partial_kf(k)\}\,.
\end{equation}
By imposing \(M\Delta k^2\text{max}\{\partial_kf(k)\}/2\ll1\), we can evaluate how large \(M\) should be so that \eqref{qfilargeM} is a good approximation of the QFI. In the limit of \(x\to0\), we find that \(\text{max}\{\partial_kf(k)\}\approx x^{-\frac{5}{2}}/5\), which implies that, to have a negligible absolute error, we require \(M\gg2\pi^2x^{-\frac{5}{2}}/5\). This indicates that the closer we are to the critical point, the larger the \(M\) needed to ensure the validity of \eqref{qfilargeM}. 

The same reasoning can be applied to the average number of photons in the ground state \eqref{photonMM}. However, the resulting integral cannot be computed analytically as we did with the QFI. An approximated analytical solution in the limit \(x\to0\) is
\begin{equation}
    N\approx M\left[\frac{1}{2\pi}\log\left(\frac{8}{x}\right)-\frac{1}{2}\right]\,.
\end{equation}
Notice that in this regime, the local number of photons \(N_{\rm loc}=N/M\) is independent of the number of modes but depends only on the proximity to the critical point. In this case, the parameter \(x\) can be expressed in terms of \(N_{\rm loc}\) as \(x\approx 8e^{-(1+2N_{\rm loc})\pi}\). Furthermore, the protocol duration time is approximately \(T\approx(\eta\lambda_0)^{-1}\approx(\eta\omega\sqrt{2x})^{-1}\), which in terms of \(N_{\rm loc}\) becomes \(T\approx e^{\left(\frac{1}{2}+N_{\rm loc}\right)\pi}/4\eta\omega\). Using these results, in terms of the involved resources we express the QFI as
\begin{equation}\label{QFIinfiniteM}
    \mathcal{I}\sim \frac{e^{\frac{\pi}{2}}}{16}\eta^2 M T^2e^{\pi N_{\rm loc}}\,. 
\end{equation}
By looking at \eqref{QFIinfiniteM}, one might conclude that the optimal operating regime of the critical chain is the continuous limit, given that the QFI scales exponentially with the photon number resource.  However, the validity of this expression is related to \eqref{errore}. As mentioned above, approximating the discrete sums defining the multi-mode QFI with definite integrals is a good approximation only if \(M\gg2\pi^2x^{-\frac{5}{2}}/5\). Considering the expression of \(x\) in terms of \(N_{\rm loc}\), we find that \(M\) must satisfy \(M\gg Ce^{5\pi N_{\rm loc}}\), with \(C=\pi^2e^{\frac{5\pi}{2}}/320\sqrt{2}\). This reveals that even for a low number of photons, the required \(M\) to ensure a negligible error is extremely large. In Fig.~\ref{QFIfigures}(c), we plot the ratio \(\log_{10}\left(\mathcal{I}/T^2\right)\) as a function of \(N_{\rm loc}\), and compare the exact optimized ratio obtained numerically from \eqref{QFIO} with the quadratic scaling of \eqref{QFIOscaling} and the exponential scaling of \eqref{QFIinfiniteM}. As expected, the QFI goes exponentially with \(N_{\rm loc}\) only when \(N_{\rm loc}=\mathcal{O}(1)\). Indeed, as soon as \(N_{\rm loc}\) becomes comparable with \(M\), the QFI is well approximated by the quadratic regime of \eqref{QFIOscaling}. The exponential in Eq.~\eqref{QFIinfiniteM} is then only an apparent super-Heisenberg scaling.
\section{Performance analyses}
\label{SV}
\noindent We can now compare the performance of the critical resonator chain with that of an ensemble of \(M\) independent single-mode critical resonators. In the following, we first thoroughly analyze the collective precision enhancement of the array of critical resonators in the relevant asymptotic regimes. Then, we assess the validity of the Gaussian approximation of the critical models and provide an estimate of the saturation of the QFI due to finite Kerr nonlinearity. Finally, we discuss the effects of coupling with a Markovian  bath, identifying the regimes in which the collective advantage can be achieved in the dissipative setting.
\begin{figure}[t!]
  \includegraphics[width=0.48
\textwidth]{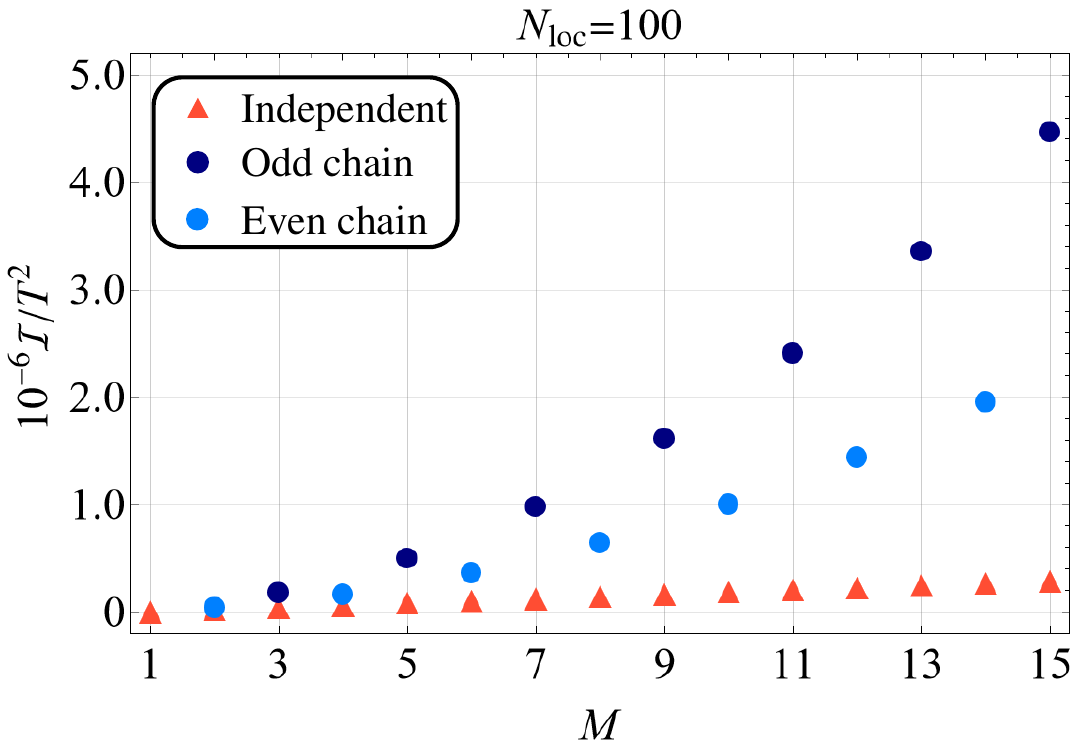}
\centering
\caption{\textbf{Quantum Fisher information: comparison between independent ensemble and critical chain.}  We plot the ratio \(\mathcal{I}/T^2\) as a function of the number of modes \(M\) in the case of independent modes \eqref{Ism} (triangles) and coupled modes \eqref{QFIE}-\eqref{QFIO} (circles). While the QFI scales linearly with \(M\) for decoupled modes, interactions between the bosonic modes enhance the QFI, leading to a quadratic scaling with respect to \(M\). To obtain the plot, for each value of \(M\) we have optimized the QFI under the constraint \(N_{\rm loc}=100\), and we have set \(\omega=1\).   }
\label{DvsC}
\end{figure}
\subsection{Asymptotic scaling}\label{SVA} We first examine the case of an ensemble of \(M\) uncoupled critical resonators. This scenario corresponds to applying the CQS protocol with a single-mode system \(M\) times, independently. For independent measurements, the QFI is additive, meaning that overall \(\mathcal{I}_{\rm ind}=M\mathcal{I}_{\rm sm}\). For an array of independent critical sensors, from \eqref{ISMscaling} we obtain 
\begin{equation}
\mathcal{I}_{\rm ind}\sim 2 \eta^ 2 T^2 M N_{\rm loc}^2 = 2 \eta^ 2 T^2\frac{N^2}{M}\,.
\end{equation}
Conversely, for a critical chain of \(M\) interacting resonators, the optimal QFI scales as 
\begin{eqnarray}
\label{static_scalingE}
\mathcal{I}_E &\sim&  \textcolor{white}{2}\eta^2 T^2 M^2 N_{\rm loc}^2 =  \textcolor{white}{2}\eta^2 T^2  N^2\,, \\\label{static_scaling}
\mathcal{I}_O &\sim& 2 \eta^2 T^2 M^2 N_{\rm loc}^2 = 2 \eta^2 T^2 N^2\,,
\end{eqnarray}
for an even and odd \(M\), respectively. The critical resonator chain presents a collective enhancement in sensing precision, resulting in a quadratic scaling of the QFI with respect to $M$. In Fig.~\ref{DvsC}, we show the growth of the QFI with respect to $M$ for fixed values of $N_{\rm loc}$, comparing even and odd critical chains with an equivalent ensemble of uncoupled resonators. The plot is numerically computed from Eq.~\eqref{Ism} and Eqs.~\eqref{QFIE}-\eqref{QFIO}.

This collective enhancement can be understood by analyzing the photon distribution in the modes $k$ in the reciprocal space. Let us focus first on a chain with an odd $M$. In proximity of the critical point the photons are strongly confined in the single mode $a_0${\newtext{, see Fig.~\ref{PhotonNumber}}}. As can be seen in Eq.~\eqref{HKdiagonal}, this mode is effectively the closest to its critical point. Given that the QFI is nonlinear with respect to the photon number, it is more favorable to concentrate all the photons into the most critical mode, rather than distributing them uniformly. This also explains the origin of the $1/2$ prefactor in the scaling of the QFI for an even chain: when $M$ is even the photons are confined into the two modes $\{a_0,a_\pi \}$, so the collective enhancement is achieved for two uniformly populated modes. In the following, we demonstrate that confining all photons into a single critical mode is advantageous also when considering saturation.
\subsection{Saturation for finite nonlinearity}\label{SVB} Let us now assess the validity of the Gaussian approximation, which is strictly valid for a vanishing nonlinearity (\(\chi=0\)) [see Eqs.~\eqref{Kerr} and \eqref{Hm_complete}]. In any physical implementation, nonlinear effects will inevitably play a role at high energies, and non-Gaussian properties emerge~\cite{Calvanese24}. For finite values of \(\chi\), as the critical point is approached, the Kerr term induces saturation in the photon number and, consequently, in the QFI. For a single critical sensor, the maximum achievable QFI for finite $\chi$ can be evaluated numerically~\cite{DiCandia2023}. However, numerical simulations of the full quantum model for large chains are beyond the capabilities of classical computers.
We then take a perturbative approach to assess the saturation point of the QFI as a function of the ratio between the frequency $\omega$ and the nonlinearity $\chi$. For the single-mode system, we applied perturbation theory directly on the QFI, the details can be found in Appendix \ref{APPC}. In this case, the validity region of the Gaussian approximation is given by
\begin{equation}\label{singolomodoboundfotoni}
    N_{\rm loc}\lesssim\sqrt[3]{\frac{\omega}{132\chi}}\,,
\end{equation}
where \(N_{\rm loc}\) is the number of photons in the unperturbed single-mode ground state. Using the bound in \eqref{singolomodoboundfotoni}, we can establish an upper limit also for the QFI in relation to the physical parameters \(\omega\) and \(\chi\). Indeed, recalling that, for a single-mode critical resonator \(T\approx 2N_{\rm loc}/\eta\omega\), the scaling of the QFI can be written solely in terms of the number of photons as \(\mathcal{I_{\rm sm}}\sim8N_{\rm loc}^4/\omega^2\). Thus, for an ensemble of \(M\) independent critical resonators, the saturation of the QFI is determined by the bound
\begin{equation}\label{singolomodoboundQFI}
    \mathcal{I}_{\rm ind}\sim\frac{8MN_{\rm loc}^4}{\omega^2}\lesssim \frac{M}{100}\sqrt[3]{\frac{1}{\omega^2\chi^4}} 
    = \frac{M}{ 100 \omega^2}\left(\frac{\omega}{\chi} \right)^{4/3} \,.
\end{equation}

For the critical chain, we focus exclusively on the case where \(M\) is odd, as it yields the most favorable QFI scaling. Applying perturbation theory in this case is much more complex, as the Kerr terms in \eqref{Hm_complete} are spatially local, but highly non-local in the reciprocal space. However, close to the criticality (when saturation effects are relevant), only a single critical mode is effectively populated. In 
Appendix \ref{APPD}, we show that when $x\to 0$ the Hamiltonian of the critical chain can be mapped onto a single critical mode, with a rescaled nonlinearity $\chi/M$. In other words, for what concerns the critical scaling, the nonlinearity of the interacting chain is effectively diluted. Under single-mode approximation, we can calculate perturbations to the QFI in spite of the nonlocality of the Kerr terms in the reciprocal space. In addition, we confirm also the validity of this single-mode approximation by evaluating perturbative corrections to the energy gap between the ground and first-excited states in the multi-mode case. This perturbative analysis reveals that, for the critical chain, the Gaussian approximation holds true if
\begin{equation}
    N\lesssim\sqrt[3]{\frac{\omega M}{132\chi}}\,,
\end{equation}
where now \(N=MN_{\rm loc}\) is the number of photons in the multi-mode ground state. It follows that, given \(T\approx 2MN_{\rm loc}/\eta\omega\), the multi-mode QFI \eqref{QFIOscaling} is asymptotically upper-bounded by
\begin{equation}\label{multimodoboundQFI}
    \mathcal{I}\sim\frac{8M^4N_{\rm loc}^4}{\omega^2}\lesssim \frac{1}{100}\sqrt[3]{\frac{M^4}{\omega^2\chi^4}}
    = \frac{1}{ 100 \omega^2}\left(\frac{M\omega}{\chi} \right)^{4/3}
    \,.
\end{equation}
Note that the asymptotic bound \eqref{multimodoboundQFI} is \(\sqrt[3]{M}\) times larger than the single-mode bound \eqref{singolomodoboundQFI}. This shows that the collective enhancement of the critical chain is not canceled by saturation effects. On the contrary, the critical chain presents a scaling advantage also with respect to the ratio $\omega/\chi$.
\subsection{Driven-dissipative case }\label{SVC}
Let us now consider an open-quantum-system setting. Notice that an in-depth analysis would require considering a microscopic derivation of the bath properties, which strongly depend on the specific physical platform and the specific readout mechanism. For example, in a chain of solid-state resonators only one or two sites would be coupled with input-output transmission lines. Furthermore, the multi-mode nature of the output field should be taken into account to assess the precision that can be achieved with continuous-measurement schemes~\cite{Ilias2022,Yang2022}.
Here, our goal is to show the conceptual differences between the performances achievable with unitary dynamics and with an open quantum system. Accordingly, we will assume that each resonator is weakly coupled to independent Markovian environments at zero temperature. In this case, the system time evolution can be written in terms of the Lindblad equation
\begin{equation}
\label{Lindblad}
\frac{\partial \rho}{\partial t} = -i \left[ H, \rho \right] + \Gamma\sum_j \mathcal D_j [\rho]\,,
\end{equation}
where $\Gamma$ is the decay rate, and we defined
\begin{equation}
\mathcal D_j [\rho] = 2 a_j \rho a^\dagger_j - a_j^\dagger a_ j \rho - \rho a_j^\dagger a_ j\,.
\end{equation}
Also in the dissipative case, we can switch to the Fourier space. It is easy to show that $\sum_j \mathcal D_j [\rho] = \sum_k \mathcal D_k [\rho]$, which means that the dissipation acts locally on the operators defined in the momentum space. For the sake of simplicity, we will restrict the analysis to odd resonator chains and to the critical region. As previously discussed, when $x\to 0$ only a single collective mode gives a relevant contribution to the QFI. Consequently, we will consider the dissipative evolution \eqref{Lindblad} of the $k=0$ mode defined  in Eq.~\eqref{Hmk}. This simplification does not affect the main results, as including all modes can only increase the estimation precision.

The single-mode case has been thoroughly analyzed \cite{alushi2024optimality}. Here, we are going to use those results to analyze the differences in the multi-mode case. So far in this work, we have considered \emph{static} CQS protocols, where an adiabatic sweep is used to push the system in proximity of the critical point. In the dissipative case, static CQS protocols consists in exploiting the susceptibility of the system-steady state, achieved after a long evolution time determined by the inverse of the smallest Liouvillian eigenvalue. By computing the QFI over the steady-state manifold, for an odd chain we find $\mathcal{I}_O \sim N^2/\Gamma^2 = M^2 N_{\rm loc}^2/\Gamma^2$. Whenever time is not considered as a resource, a quadratic enhancement in $M$ can be found in static dissipative protocols. However, the time required to reach the steady-state scales as $T_{\rm ss} \sim MN_{\rm loc}/\Gamma$, meaning that $\mathcal{I}_O \sim M N_{\rm loc}T_{\rm ss}/\Gamma$. Consequently, when time is a resource no collective advantage is found over the use of $M$ independent single-mode sensors (recall that \(\mathcal{I}_{\rm ind}=M\mathcal{I}_{\rm sm}\)).

\begin{figure}[t!]
  \includegraphics[width=0.48
\textwidth]{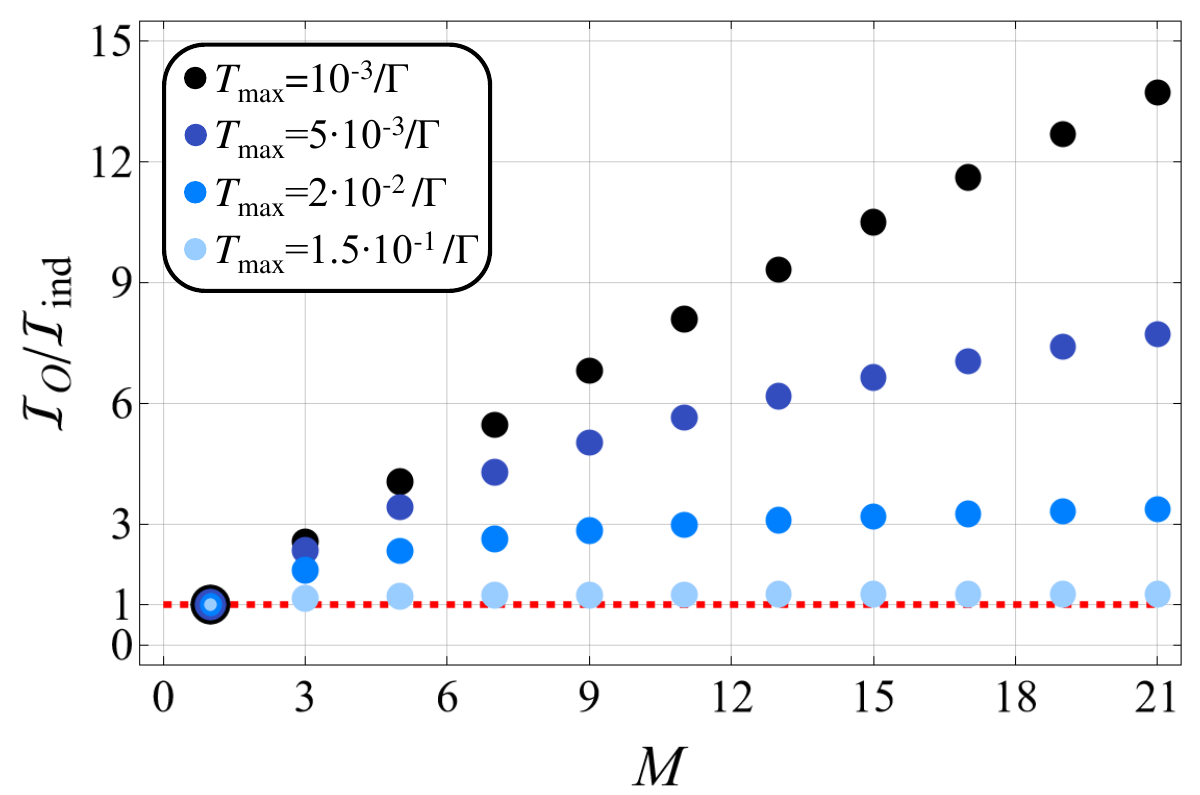}
\centering
\caption{{\textbf{Collective enhancement in the driven-dissipative case.}} Plot of the instantaneous quantum Fisher information ratio \(\mathcal{I}_O/\mathcal{I}_{\rm ind}\) as a function of the number of modes \(M\). We plot the ratio at different time snapshots \(T_{\rm max}\), considering a dynamical protocol. Specifically, for a time \(T_{\rm max}\lesssim3/(4M N_{\rm loc} \Gamma)\) the system dynamics is still unitary and we observe a collective enhancement in \(M\) as \(\mathcal{I}_O\sim M^\alpha\), with \(1<\alpha\leq2\). When \(T_{\rm max}> 3/(4M N_{\rm loc} \Gamma)\), the scaling coefficient becomes \(\alpha\approx1\). However,  it is still possible to have a constant advantage using the coupled-resonator chain since \(\mathcal{I}_O/\mathcal{I}_{\rm ind}>1\). This advantage is washed out when the time is long enough to make the advantage due to short-time unitary evolution negligible. To obtain the plot, we considered \(\omega=\sqrt{MN_{\rm loc}/T^2_{\rm max}}\Gamma\), which is optimal in the unitary regime \cite{alushi2024optimality}. In addition, we set \(N_{\rm loc}=10\), and we work close to the criticality (regime in which the odd multi-mode array can be mapped onto a single critical mode). Specifically, we choose \(\epsilon\) in order to satisfy the constraint \(N(t=T_{\rm max})=MN_{\rm loc}\). In red (dashed line) we highlight the reference case \(\mathcal{I}_O/\mathcal{I}_{\rm ind}=1\), which corresponds to the absence of any collective advantage.}  
\label{DissipativeResistence}
\end{figure}

We now show that a quantum advantage in $M$ can be found in the dissipative case even for finite evolution times, when a \emph{dynamical} CQS protocol is used. In this case, the scaling of the QFI depends on the protocol duration time itself, so we fix a maximum time $T_{\rm max}$ in which the protocol can be performed. In this way, time is treated as a resource on the same footing of the local photon number, for which we impose a maximum value $N_{\rm loc}$. We assume the system is initialized in the vacuum $\rho(0) = |0\rangle\langle0|$. Then, we evaluate the instantaneous QFI on the state $\rho(t)$ evolved according to Eq.~\eqref{Lindblad}. This evaluation can be done analytically at arbitrary time~\cite{alushi2024optimality}. However, the expression of the instantaneous QFI is too complex to show explicitly. In the following, we present approximate solutions for the unitary regime, and we analyze the general behavior in Fig.~\ref{DissipativeResistence}. For short times, we can expand the QFI in series up to the first order in $\Gamma$ as
\begin{equation}
\label{QFIexpanded}
\mathcal{I}_O \sim \frac{8}{9} T_{\rm max}^2 N^2 - 
\frac{32}{27} T_{\rm max}^3 N^3 \Gamma + o\left(\Gamma^2 \right)\,. 
\end{equation}
For sufficiently short times, we recover the quadratic scaling in both time and photon number observed in the unitary case \eqref{static_scaling}. Indeed, from Eq.~\eqref{QFIexpanded} we identify the regime in which the quantum collective advantage of the coupled-resonator chain is present also in the dissipative case:
\begin{equation}
\label{dynamical_scaling}
\mathcal{I}_O \sim \frac{8}{9} T_{\rm max}^2 M^2 N_{\rm loc}^2 \quad 
\text{for} \quad T_{\rm max}\ll \frac{3}{4 M N_{\rm loc} \Gamma}\,. 
\end{equation}
It is important to notice that the time in which the quantum advantage is not spoiled by dissipation is bounded by the resources themselves, in agreement with fundamental bounds~\cite{Gorecki_bounds, demkowicz2017adaptive}.
Comparing Eqs.~\eqref{dynamical_scaling} and ~\eqref{static_scaling}, one can also see the constant-factor advantage of dynamical over static CQS protocols~\cite{Chu2021}, due to the  adiabatic parameter $\eta\ll1$. Finally, in Fig.~\ref{DissipativeResistence} we analyze the 
 transition from the unitary to the dissipative regime, for different number of modes $M$. For sufficiently short $T_{\rm max}$, the dynamics is effectively unitary (black dots), so that $\mathcal{I}_O\sim M^2$ while $\mathcal{I}_{\rm ind} \sim M$. Notice that, as $T_{\rm max}$ is increased, even though the quadratic advantage in $M$ fades there is a wide transition during which $\mathcal{I}_O\sim M^\alpha$ with $\alpha>1$. This behavior is reminiscent with what has been predicted for uncertainties on the system parameters~\cite{Mihailescu24}. Interestingly, from Fig.~\ref{DissipativeResistence} we see that a constant-factor advantage in the collective case can still be observed for large $M$.

\section{DISCUSSION AND PERSPECTIVES}
\label{SVD}
We have demonstrated that a critical quantum sensor based on a chain of \(M\) parametrically coupled Kerr resonators exhibits a quadratic enhancement of the QFI with respect to both photon and resonator numbers. Beyond asymptotic scalings, the coupled chain offers practical advantages in terms of physical parameters. This collective quantum advantage in critical quantum sensing is significant even for a small number of resonators. We have analyzed a variety of parameter regimes, and 
we summarize the QFI scaling for the most interesting cases relevant to quantum sensing applications in Table \ref{Table1}. 

\begin{table}[h!]
\centering
\begin{tabular}{ | c ||c|| c| } 
 \hline
 \multicolumn{3}{|c|}{ Scaling of the QFI} \rule[-2ex]{0pt}{3ex}\\
 \hline \hline
& Independent {Sensors} & Coupled chain \rule[-2ex]{0pt}{3ex}\\
 \hline
$N_{\rm loc}$ scaling 
& $2\eta^ 2 T^2 M N_{\rm loc}^2$ 
& $2 \eta^2 T^2 M^2 N_{\rm loc}^2 $ \rule[-2ex]{0pt}{5ex}\\
\hline
$N$ scaling 
& $2 \eta^ 2 T^2\frac{N^2}{M}$  
& $2 \eta^2 T^2  N^2$ \rule[-2ex]{0pt}{5ex}\\
\hline
Saturation 
& $\frac{M}{ 100 \omega^2}\left(\frac{\omega}{\chi} \right)^{4/3}$ 
& $\frac{1}{ 100 \omega^2}\left(\frac{M\omega}{\chi} \right)^{4/3}$ \rule[-2ex]{0pt}{6ex} \\
\hline
\end{tabular}
\caption{\textbf{Quantum Fisher information scaling.} The table shows the scaling of the quantum Fisher information in the case of independent modes and coupled modes. }\label{Table1}
\end{table}
\noindent This work paves the way to the exploration of collective quantum advantage in critical quantum sensing, both at experimental and theoretical levels. The considered class of finite-component phase transitions has already been experimentally observed using atomic~\cite{cai2021observation},  polaritonic~\cite{DelteilNatMat19,Zejian2022} and circuit-QED ~\cite{Fink2017,Brooks2021,beaulieu2023observation,chen2023quantum,Sett24} devices. Parametric resonators with extremely low static nonlinearity can be implemented using high kinetic inductance cavity arrays~\cite{jouanny2024}. More generally, chains of nonlinear coupled-resonator chains can be implemented using a broad variety of solid-state platforms, including semiconducting~\cite{RevModPhys.85.299}, opto- and electro-mechanical systems~\cite{barzanjeh2022optomechanics,Slim_2024} and superconducting quantum circuits~\cite{GU20171}, among others. Further theoretical work is needed to properly support the development of collective CQS protocols. The driven-dissipative case should be addressed considering platform-specific couplings with the environment, which will also depend on the chosen readout mechanism. The need for adiabatic processes, responsible for the $\eta<1$ prefactor in the scaling of the QFI, can be lifted by designing dynamical protocols~\cite{Chu2021,Garbe2022}. Continuous-measurement schemes~\cite{Ilias2022,Yang2022} can be designed to optimize information retrieval from the output signals. Finally, topological properties have been shown~\cite{mcdonald2020exponentially} to enhance the sensing precision in sensors based on non-Hermitian lattice dynamics. The inclusion of topological properties in critical chains could potentially enhance the precision and resilience to parameter fluctuations of collective CQS protocols. 
Besides quantum sensing and metrology, this work presents a phenomenological study of the ground-state properties of an unconventional quantum many-body model, which is critical both at the local and global levels. We analyzed the regime of weak nonlinearity before symmetry breaking occurs. Numerical simulations of the full quantum model are challenging due to large photon numbers and long-range correlations. Phase-space methods appear to be the most suitable for analyzing lattices of critical resonators with weak but finite nonlinearity, where the ground state of the system is approximately Gaussian.
\begin{acknowledgments}
U.A. and R.D. acknowledge financial support from the Academy of Finland, grants no. 353832 and 349199.
A.C., V.B. and S.F. acknowledge financial support from PNRR MUR project PE0000023-NQSTI financed by the European Union – Next Generation EU. 
V.B. acknowledges support from PON Ricerca e Innovazione 2014-2020 FESR 7 FSC - Project ARS01 00734 QUANCOM.
\end{acknowledgments}

\appendix
\section{Local number of photons}\label{AppendixPhotons}
In this section, we provide details on the calculation of the local number of photons in the critical resonator chain. The local number of photons in the ground state is given by
\begin{equation}\label{Aphoton}
    N_{\rm loc}^j=\bra{g} a_j^\dag a_j\ket{g}\,,
\end{equation}
where \(\ket{g}\) is defined in \eqref{gk}. We express the operators \(a_j\) in the reciprocal space as
\begin{align}
   a_j&=\frac{1}{\sqrt{M}}\sum_{k}a_k e^{-ikj}\,,\\\quad
    a^\dag_j&=\frac{1}{\sqrt{M}}\sum_{k}a^\dag_k e^{ikj}\,,
\end{align}
and rewrite the expectation value in \eqref{Aphoton}:
\begin{equation}
    N_{\rm loc}^j=\bra{g}\frac{1}{M}\sum_{k,k'}a^\dag_k a_{k'} e^{-i(k'-k)j}\ket{g}\,.
\end{equation}
Thus, to obtain \(N_{\rm loc}\), we have now to compute the expectation values \(\bra{g}a^\dag_k a_{k'}\ket{g}\), knowing that the ground state is the multi-mode squeezed vacuum state of Eq.~\eqref{gk}. Let us consider first the case \((k,\,k')\neq0\) and \((k,\,k')\neq\pi\). Then, the expectation value becomes
\begin{align}
 \bra{g}a^\dag_k a_{k'}\ket{g}&=\bra{0}\bigotimes_{k''>0}S^\dag_{k'',-k''}a_k^\dag a_{k'}\bigotimes_{k''>0}S_{k'',-k''}\ket{0}=\nonumber\\\quad
 &=\bra{0}S^\dag_{k,-k}a^\dag_k a_k S_{k,-k}\ket{0}=\nonumber\\\quad&=\bra{g}a^\dag_k a_k \ket{g}\,,
\end{align}
which is valid since, in the case in which \(k\neq k'\neq k''\), all the operators commute, and the expectation values are null. Following the same reasoning, if \(k\) or \(k'\) are equal to \(0\) or \(\pi\), the expectation value is always zero unless \(k=k'\). It follows that
\begin{equation}\label{Aphotonk}
    N_{\rm loc}^j=\bra{g}\frac{1}{M}\sum_k a^\dag_k a_k\ket{g}=N_{\rm loc}\,,
\end{equation}
which does not depend on the particular site location \(j\). The photons in the system are uniformly distributed across each cavity. We can further simplify \eqref{Aphotonk} by expanding the discrete sum over \(k>0\):
\begin{align}
    N_{\rm loc}&=\frac{1}{M}\bra{g}a^\dag_0a_0+\delta_{N,2m}a^\dag_{\pi} a_{\pi}\ket{g}+\nonumber\\\quad&+\frac{1}{M}\bra{g}\sum_{k>0}\left(a^\dag_ka_k+a^\dag_{-k}a_{-k}   
   \right)\ket{g}\,.
\end{align}
From Eq.~\eqref{gk}, the photon number in each collective mode \(a_{k}\) is \(N_k=\sinh^2\left(|\xi_k|\right)=s_k^2\), where \(s_k\) is defined as in \eqref{s} with the substitution \(\epsilon\to\epsilon\cos(k)\). Finally, since \(s^2_k=s^2_{-k}\), we write the local number of photons as
\begin{equation}
    N_{\rm loc}=\frac{1}{M}\sum_k N_k=\frac{1}{M}\left(s_0^2+\delta_{M,2m}s_{\pi}^2+\sum_{k>0}2s_k^2\right)\,.
\end{equation}

\section{Series expansions  }\label{AppendixSeries}
In this section, we present the mathematical details concerning the trigonometric series that appear when computing the QFI and \(N_{\rm loc}\) in the asymptotic limit \(x\to0\). Let us first consider the QFI when \(M\) is odd. Expanding the QFI \eqref{QFIO} up to the first order in \(x\) yields
\begin{align}
    \mathcal{I}_{\rm O}&=\frac{1}{8\omega^2}\left(\frac{1}{x^2}-\frac{1}{x}-\frac{1}{4}\right)+\nonumber\\\quad&+\frac{1}{\omega^2}\sum_{n=1}^{\frac{M-1}{2}}\left[\csc^4\left(\frac{2\pi n}{M}\right)-\csc^2\left(\frac{2\pi n}{M}\right)\right]+o(x)\,.
\end{align}
A closed analytical expression for the QFI in this limit can be obtained using the following identities on trigonometric series \cite{grabner2007secant}:
\begin{align}
\sum_{n=1}^{\frac{M-1}{2}}\csc^2\left(\frac{2n\pi}{M}\right)&=\frac{M^2-1}{6}\,,\label{C1}\\\quad\sum_{n=1}^{\frac{M-1}{2}}\csc^4\left(\frac{2n\pi}{M}\right)&=\frac{M^4+10M^2-11}{90}\label{C2}\,.
\end{align}
Thus, in our case we have
\begin{align}
\sum_{n=1}^{\frac{M-1}{2}}\left[\csc^4\left(\frac{2n\pi}{M}\right)-\csc^2\left(\frac{2n\pi}{M}\right)\right]=\frac{M^4-5M^2+4}{90}\,,
\end{align}
which then leads to \eqref{QFIOasy}.
On the other hand, when \(M\) is even, expanding the QFI \eqref{QFIE} leads to
\begin{align}
    \mathcal{I}_{\rm E}&=\frac{1}{4\omega^2}\left(\frac{1}{x^2}-\frac{1}{x}-\frac{1}{4}\right)+\nonumber\\\quad&+\frac{1}{\omega^2}\sum_{n=1}^{\frac{M}{2}-1}\left[\csc^4\left(\frac{2\pi n}{M}\right)-\csc^2\left(\frac{2\pi n}{M}\right)\right]+o(x)\,.
\end{align}
From the symmetry properties of the cosecant functions and their periodicity, it follows
\begin{equation}
\sum_{n=1}^{\frac{M}{2}-1}\left[\csc^4\left(\frac{2n\pi}{M}\right)-\csc^2\left(\frac{2n\pi}{M}\right)\right]=\frac{M^4-20M^2+64}{720}\,,
\end{equation}
which then leads to \eqref{QFIEasy}.

Let us now focus on the average number of photons. For an even \(M\), expanding \eqref{photonMM} up to the first order in \(x\) yields
\begin{equation}
    N=\frac{1}{\sqrt{2x}}-\frac{M}{2}+\sum_{n=1}^{\frac{M}{2}-1}\csc\left(\frac{2n\pi}{M}\right)+o(\sqrt{x})\,.
\end{equation}
The summation of the cosecant functions can be approximated as ~\cite{blagouchine2024finite}
\begin{equation}
    \sum_{n=1}^{\frac{M}{2}-1}\csc\left(\frac{2n\pi}{M}\right)=\frac{M}{\pi}\left[\gamma+\log\left(\frac{M}{\pi}\right)\right]+\mathcal{O}\left(\frac{1}{M}\right)\,,
\end{equation}
where \(\gamma\) is the Euler-Mascheroni constant. Thus, we obtain Eq.~\eqref{NE}. Conversely, for an odd \(M\), expanding in series \eqref{photonMM} results in
\begin{equation}
    N=\frac{1}{\sqrt{8x}}-\frac{M}{2}+\sum_{n=1}^{\frac{M-1}{2}}\csc\left(\frac{2n\pi}{M}\right)+o(\sqrt{x})\,. 
\end{equation}
Due to the symmetry properties of the cosecant function and its periodicity, we have
\begin{equation}
\sum_{n=1}^{\frac{M-1}{2}}\csc\left(\frac{2n\pi}{M}\right)=\frac{M}{\pi}\left[\gamma+\log\left(\frac{2M}{\pi}\right)\right]+\mathcal{O}\left(\frac{1}{M}\right)\,,
\end{equation}
from which we retrieve Eq.~\eqref{NO}.

\section{Finite non-linearity: single-mode}\label{APPC}
In this section, we go through the details of the perturbation theory applied to assess the validity of the Gaussian approximation. \\
The squeezing Hamiltonian \eqref{Hsm} is an approximation of the general model:
\begin{equation}
    H=\omega a^\dag a+\frac{\epsilon}{2}\left(a^2+a^{\dag2}\right)+\chi a^\dag a^\dag a a\,,
\end{equation}
with \(\chi\) representing the Kerr non-linearity. The Gaussian approximation is obtained by setting \(\chi=0\). However, when the number of photons in the system state increases beyond a certain limit, dependent on \(\chi\) and \(\omega\), the nonlinear term becomes significant and cannot be neglected. We take a perturbative approach to identify an upper bound on the allowed number of photons, and, consequently, estimate the saturation of the QFI.

Let us start by analysing the energy gap \(\Delta E\) between the ground and the first excited state. Up to the first order in \(\chi\), it is defined as 
\begin{equation}\label{energygapperturbation}
\Delta E=\Delta E^{(0)}+\chi\Delta E^{(1)}+o\left(\chi^2\right)\,, 
\end{equation}
where \(\Delta E^{(0)}=\lambda=\sqrt{\omega^2-\epsilon^2}\) and \(\Delta E^{(1)}=\bra{e}H_{\rm NL}\ket{e}-\bra{g}H_{\rm NL}\ket{g}\). Here, \(\ket{g}=S(\xi)\ket{0}\) is the ground state, \(\ket{e}=S(\xi)\ket{1}\) is the first excited state and \(H_{\rm NL}=a^\dag a^\dag a a \). We recall that the squeezing parameter is defined as \(\xi=\ln\left(s+\sqrt{s^2+1}\right)\), with \(s\) given by \eqref{s}. Since \(\ket{g}\) and \(\ket{e}\) are single-mode squeezed Fock states, \(\Delta E^{(1)}\) can be easily computed as
\begin{align}
    \Delta E^{(1)}&=\bra{e}H_{\rm NL}\ket{e}-\bra{g}H_{\rm NL}\ket{g}=\nonumber\\\quad&=\left(9N+15N^2\right)-\left(N+3N^2\right)=\nonumber\\\quad&=8N+12N^2\,,
\end{align}
with \(N=\sinh^2\left(|\xi|\right)\) the number of photons in the ground state. In the Gaussian approximation, we have \(\Delta E=\Delta E^{(0)}\). Thus, when \(\chi>0\), it must hold that
\begin{equation}
    \frac{\chi \Delta E^{(1)}}{\Delta E^{(0)}}=\frac{\chi\left(8N+12N^2\right)}{\lambda}\ll1\,.
\end{equation}
By expressing \(\lambda\) in terms of \(N\) as \(\lambda=\omega/(1+2N)\), we obtain an estimate of the desired upper bound on the number of photons in the ground state, i.e.,
\begin{equation}\label{smphbound}
    N\lesssim\sqrt[3]{\frac{\omega}{24\chi}}\,.
\end{equation}
This means that, as far as \(N\) is well below the upper bound \eqref{smphbound}, the Gaussian approximation is valid and the non-linearity can be safely neglected.\\
The same procedure may be applied directly to the QFI. In the following, we show that this leads to an upper bound with the same scaling in \(\omega/\chi\) as in \eqref{smphbound} but slightly tighter. To apply the perturbation theory to the QFI, we use \eqref{qfi} with the ground state defined as 
\begin{equation}
\ket{g}=|g^{(0)}\rangle+\chi|g^{(1)}\rangle+o\left(\chi^2\right)\,.
\end{equation}
Here, \(|{g}^{(0)}\rangle=S(\xi)\ket{0}\) is the unperturbed ground state, while \(|{g}^{(1)}\rangle\) is the first-order correction to the unperturbed ground state. Up to the first order in \(\chi\), the QFI can be expressed as
\begin{equation}
\mathcal{I}=\mathcal{I}^{(0)}+\chi \mathcal{I}^{(1)}+o\left(\chi^2\right)\,,
\end{equation}
where \(\mathcal{I}^{(0)}\) is the unperturbed QFI defined in \eqref{Ism}, while \(\mathcal{I}^{(1)}=8\Re\left[\bra{\partial_\omega g^{(0)}}\ket{\partial_\omega g^{(1)}}\right]\) is the first order correction. Perturbation theory allows us to easily compute \(|g^{(1)}\rangle\). Indeed, in our case 
\begin{equation}\label{g1formula}
|g^{(1)}\rangle=\sum_{j=1}^{4}\frac{\langle j^{(0)}| H_{\rm NL}|g^{(0)}\rangle}{-j\lambda}|j^{(0)}\rangle\,,
\end{equation}
with \(|j^{(0)}\rangle=S(\xi)\ket{j}\). The sum is performed for \(j\in[1,\,4]\), since when \(j>4\) all the expectation values are null due to the orthogonality of the Fock states. The transition elements in \eqref{g1formula} can be then analytically evaluated as \(|j^{(0)}\rangle\) are single-mode squeezed Fock states. The first order correction to the ground state is then
\begin{equation}
|g^{(1)}\rangle=v_{\omega,\epsilon}S(\xi)\ket{2}+h_{\omega,\epsilon}S(\xi)\ket{4}\,.
\end{equation}
The amplitude probability functions \(v_{\omega,\epsilon}\) and \(h_{\omega,\epsilon}\) are defined as
\begin{align}
v_{\omega,\epsilon}&=\frac{(1+6N)\sqrt{2(N+N^2)}}{2\lambda}\,,\nonumber\\\quad
h_{\omega,\epsilon}&=-\frac{\sqrt{24}\left(N+N^2\right)}{4\lambda}\,,
\end{align}
and depend on  \(\omega\) and \(\epsilon\) according to \(N=N(\omega,\epsilon)\) and  \(\lambda=\lambda(\omega,\epsilon)\). If we now rewrite the squeezing operator as \(S(\xi)=e^{f_{\omega,\epsilon}O}\), where \(f_{\omega,\epsilon}=|\xi|/2\) and \(O=a^2-a^{\dag2}\), the derivatives with respect to the parameter \(\omega\) read
\begin{align}\label{derivateparziali}
    |\partial_\omega g^{(0)}\rangle&=f'_{\omega,\epsilon}O|g^{(0)}\rangle\,,\nonumber\\\quad
    |\partial_\omega g^{(1)}\rangle&=\left(v'_{\omega,\epsilon}+v_{\omega,\epsilon}f'_{\omega,\epsilon}O\right)|2^{(0)}\rangle+\nonumber\\\quad&+\left(h'_{\omega,\epsilon}+h_{\omega,\epsilon}f'_{\omega,\epsilon}O\right)|4^{(0)}\rangle\,,
\end{align}
where the apex stands for \(\partial_\omega\). Finally, using \eqref{derivateparziali}, we are able to compute the first-order correction to the QFI:
\begin{align}
\mathcal{I}^{(1)}&=8\langle\partial_\omega g^{(0)}|\partial_\omega g^{(1)}\rangle=\nonumber\\\quad&=-8\sqrt{2}f'_{\omega,\epsilon}\left(v'_{\omega,\epsilon}+\sqrt{12}f'_{\omega,\epsilon}h_{\omega,\epsilon}\right)\,.
\end{align}
From the definition \(N=s^2\), we obtain an expression for \(\epsilon\) in terms of the unperturbed ground state number of photons, i.e., \(\epsilon=2\omega\sqrt{N+N^2}/\left(1+2N\right)\). With this in mind, the leading terms of the zeroth-order QFI and its first-order correction are
\begin{align}
    \mathcal{I}^{(0)}\sim \frac{8N^4}{\omega^2}\,\,\,\,\,,\,\,\,\,\,
    \mathcal{I}^{(1)}\sim -\frac{1056 N^7}{\omega^3}\,.
\end{align}
In the Gaussian approximation, it holds that \(\mathcal{I}=\mathcal{I}^{(0)}\). Thus, when \(\chi>0\), we have to impose \(|\chi \mathcal{I}^{(1)}|\ll| \mathcal{I}^{(0)}|\), from which we obtain a new estimate for the upper bound on the number of photons in the ground state:
\begin{equation}\label{smphqfibound}
    N\lesssim\sqrt[3]{\frac{\omega}{132\chi}}\,.
\end{equation}
This bound has the same scaling in \(\omega/\chi\) as that in \eqref{smphbound}, but is tighter by roughly a factor of two. \\

\section{Finite non-linearity: resonator chain}\label{APPD}
For the multi-mode case, the Hamiltonian in \eqref{Hm} can be thought of as the Gaussian approximation of the general model
\begin{equation}\label{hamiltonianokerrmulti}
H= \sum^{M}_{j=1} \left[
\omega a^\dag_ja_j + \frac{\epsilon}{2} \left( a_ja_{j+1}+a^\dag_ja^\dag_{j+1} \right)
+ \chi a^\dag_ja^\dag_ja_j a_j
\right]\,,
\end{equation}
where \(\chi\) represents the local Kerr non-linearity. To assess the validity of the Gaussian approximation, we can apply perturbation theory, as we did for the single-mode case. In the following, we consider \(M\) odd, since this choice leads to the most favorable scaling of the QFI. Close to the criticality, the majority of the photons are then confined in the mode \(a_0\), leaving the other modes \(a_{\pm k}\) unpopulated. Thus, we approximate the unperturbed ground state defined in \eqref{gk} as
\begin{equation}\label{gkapprossimato}
    |g^{(0)}\rangle\approx S_0(\xi_0)\ket{0}\,.
\end{equation}
This allows us to map the multi-mode system to a single-mode system. The low-energy excited states can be written as 
$|j^{(0)}\rangle\approx \left(j!\right)^{-\frac{1}{2}} S_0(\xi_0)\,(a_0^{\dag})^j\ket{0}$. The single-mode system has now \(N=MN_{\rm loc}\) photons in the mode \(a_0\), where \(N_{\rm loc}\) is the local number of photons in each site \(j\). Let us now focus on the effect of the nonlinear terms, which can be written in the reciprocal space as
\begin{align}\label{multimodenonlinear}
   &\chi H_{\rm NL}=\chi\sum_j^{M}a^\dag_j a^\dag_j a_j a_j=\nonumber\\\quad&=\frac{\chi}{M^2}\sum_j^M\sum_{k_1,k_2,k_3,k_4}a^\dag_{k_1}a^\dag_{k_2}a_{k_3}a_{k_4}e^{-i(k_3+k_4-k_1-k_2)j}=\nonumber\\\quad
   &=\frac{\chi}{M}\sum_{k_1,k_2,k_3}a^\dag_{k_1}a^\dag_{k_2}a_{k_3}a_{k_1+k_2-k_3}\,.
\end{align}   
Given the structure of the low-excited states, the expected value and transition elements of the nonlinear term are nonvanishing only for $k_1 = k_2=k_3=k_4=0$.
In this single-mode approximation, the Hamiltonian~\eqref{hamiltonianokerrmulti} can be mapped onto a single critical resonator with a rescaled nonlinear term, such that
\begin{equation}
\chi H_{\rm NL} \longrightarrow \frac{\chi}{M} a_0^\dagger a_0^\dagger a_0 a_0\,.
\end{equation}
We can then perform the same calculations we did for the perturbation of the QFI in the single-mode case, but with the ground state \eqref{gkapprossimato} and a rescaled nonlinear Hamiltonian. It follows that the bounds found for the single-mode case are still valid for the multi-mode one, provided that we replace \(N\to MN_{\rm loc}\) and \(\chi\to\chi/M\). Finally, to ensure the validity of the Gaussian approximation in the multi-mode scenario, we require
\begin{equation}
    N\lesssim\sqrt[3]{\frac{\omega M}{132\chi}}\,.
\end{equation}
So far, we have assumed that the unperturbed multi-mode ground state can be written as \eqref{gkapprossimato}. This allowed us to evaluate the first-order correction to the QFI. In this section, we analytically compute the first-order correction to the multi-mode energy gap, without resorting to Eq.~\eqref{gkapprossimato}. In the limit where \(N\approx N_0\), with \(N\) the number of photons in the ground state and \(N_0\) the number of photons in the mode \(a_0\), the single-mode bound \eqref{smphbound} is retrieved, thus endorsing the validity of the single-mode approximation.  
The multi-mode energy gap between the ground and the first excited state is defined as in \eqref{energygapperturbation}, with \(\Delta E^{(0)}=\lambda_0=\sqrt{\omega^2-\epsilon^2}\) and \(\Delta E^{(1)}=E^{(1)}_e-E^{(1)}_g\), where \(E_e^{(1)}\) and \(E_g^{(1)}\) are the first order corrections to the eigenvalues of the first-excited and ground state, respectively. Let us start by computing the first-order correction to the ground state \(E_g^{(1)}\). By using \eqref{multimodenonlinear} and the relabeling \(k_1\to k'+q\), \(k_2\to k-q\), \(k_3\to k'\), we write \(E_g^{(1)}\) as
\begin{align}\label{E1ground}
    E^{(1)}_g&=\frac{1}{M}\sum_{k,k',q}\bra{g}a^\dag_{k'+q}a^\dag_{k-q}a_{k'}a_{k}\ket{g}=\nonumber\\\quad
    &=\frac{1}{M}\sum_{k,k',q}\bra{g}a^\dag_{k'+q}a_{k'}a^\dag_{k-q}a_{k}\ket{g}-N\,,
\end{align}
where we have used \([a_{k'},a^\dag_{k-q}]=\delta_{k',k-q}\) and \(MN=\sum_{k,q}\bra{g}a^\dag_ka_k\ket{g}\). Since the ground state \(\ket{g}\) is defined as in \eqref{gk}, any expectation value  \(\bra{g}a_k\ket{g}\) is null. This means that the non-zero terms in \eqref{E1ground} are those in which the ladder operators refer to the same mode in pairs, leading to the constraints: (i) \(k'+q=k'\) and \(k-q=k\), meaning \(q=0\); (ii) \(k'+q=k-q\) and \(k'=k\), meaning \(q=0\) and \(k=k'\); (iii) \(k'+q=k\), meaning \(q=k-k'\). Implementing these constraints in \eqref{E1ground} allows us to obtain a much simpler expression:
\begin{align}\label{E1groundsemplice}
    E^{(1)}_g&=\frac{1}{M}\Bigg[2\sum_{k,k'}\bra{g}a^\dag_{k'}a_{k'}a^\dag_ka_k\ket{g}+\nonumber\\\quad&-\sum_k\bra{g}a^\dag_ka_ka^\dag_k a_k\ket{g}-N\Bigg]\,.
\end{align}
On the other hand, the first excited state in the multi-mode system is
\begin{equation}
\ket{e}=S(\xi_0)\ket{1}\bigotimes_{k>0}S_{k,-k}(\xi_k)\ket{0_k}\,.
\end{equation}
Thus, the expression \eqref{E1groundsemplice} holds also for \(E_e^{(1)}\), provided that \(\ket{g}\to\ket{e}\) and \(N\to N+1+2N_0\) (it is straightforward to verify that the number of photons in \(\ket{e}\) is indeed \(N+1+2N_0\)). The first-order correction to the multi-mode energy gap is then
\begin{align}\label{DeltaEmultisemplice}
    &\Delta E^{(1)}=E^{(1)}_e-E^{(1)}_g=\nonumber\\\quad
    &=\frac{1}{M}\Bigg[2\sum_{k,k'}\left(\bra{e}a^\dag_{k'}a_{k'}a^\dag_ka_k\ket{e}-\bra{g}a^\dag_{k'}a_{k'}a^\dag_ka_k\ket{g}\right)+\nonumber\\\quad&-\sum_k\left(\bra{e}a^\dag_{k}a_{k}a^\dag_ka_k\ket{e}-\bra{g}a^\dag_{k}a_{k}a^\dag_ka_k\ket{g}\right)+\nonumber\\\quad&-1-2N_0\Bigg]\,.
\end{align}
Since the first excited state \(\ket{e}\) differs from ground state \(\ket{g}\) only in the terms related to the mode \(a_0\), the differences between expectation values in \eqref{DeltaEmultisemplice} will always give zero unless \(k=0\) or \(k'=0\). Then, we can further simplify  \(\Delta E^{(1)}\):
\begin{align}\label{DeltaEmultisupersemplice}
   \Delta E^{(1)}&=\frac{1}{M}\Bigg[4\left(1+2N_0\right)\left(N-N_0\right)+\nonumber\\\quad
&+\bra{e}a^\dag_0a_0a^\dag_0a_0\ket{e}-\bra{g}a^\dag_0a_0a^\dag_0a_0\ket{g}+\nonumber\\\quad
   &-1-2N_0\Bigg]\,.
\end{align}
The remaining expectation values in \eqref{DeltaEmultisupersemplice} can be computed easily, being \(\ket{e}\) and \(\ket{g}\) single-mode squeezed Fock states for what concerns \(a_0\). We have that \(\bra{e}a^\dag_0a_0a^\dag_0a_0\ket{e}=1+12N_0+15N_0^2\) and \(\bra{g}a^\dag_0a_0a^\dag_0a_0\ket{g}=2N_0+3N_0^2\). Finally, we write
\begin{equation}
    \Delta E^{(1)}=\frac{1}{M}\left[4N+8N_0N+4N_0+4N_0^2\right]\,,
\end{equation}
and, up to the first order in \(\chi\), the multi-mode energy gap is given by
\begin{equation}
    \Delta E=\lambda_0+\frac{\chi}{M}\left[4N+8N_0N+4N_0+4N_0^2\right]+o(\chi^2)\,.
\end{equation}
Close to the critical point, we have \(N=MN_{\rm loc}\approx N_0\), so that
\begin{equation}
    \Delta E\approx\lambda_0+\frac{\chi}{M}\left[8N+12N^2\right]\,.
\end{equation}
We retrieve the same result obtained in the single-mode case in \eqref{smphbound}, with a larger number of photons \(N=MN_{\rm loc}\) and a rescaled non-linearity \(\chi/M\). 
\bibliography{bibliofinal.bib}

\begin{thebibliography}{82}%
\makeatletter
\providecommand \@ifxundefined [1]{%
 \@ifx{#1\undefined}
}%
\providecommand \@ifnum [1]{%
 \ifnum #1\expandafter \@firstoftwo
 \else \expandafter \@secondoftwo
 \fi
}%
\providecommand \@ifx [1]{%
 \ifx #1\expandafter \@firstoftwo
 \else \expandafter \@secondoftwo
 \fi
}%
\providecommand \natexlab [1]{#1}%
\providecommand \enquote  [1]{``#1''}%
\providecommand \bibnamefont  [1]{#1}%
\providecommand \bibfnamefont [1]{#1}%
\providecommand \citenamefont [1]{#1}%
\providecommand \href@noop [0]{\@secondoftwo}%
\providecommand \href [0]{\begingroup \@sanitize@url \@href}%
\providecommand \@href[1]{\@@startlink{#1}\@@href}%
\providecommand \@@href[1]{\endgroup#1\@@endlink}%
\providecommand \@sanitize@url [0]{\catcode `\\12\catcode `\$12\catcode `\&12\catcode `\#12\catcode `\^12\catcode `\_12\catcode `\%12\relax}%
\providecommand \@@startlink[1]{}%
\providecommand \@@endlink[0]{}%
\providecommand \url  [0]{\begingroup\@sanitize@url \@url }%
\providecommand \@url [1]{\endgroup\@href {#1}{\urlprefix }}%
\providecommand \urlprefix  [0]{URL }%
\providecommand \Eprint [0]{\href }%
\providecommand \doibase [0]{https://doi.org/}%
\providecommand \selectlanguage [0]{\@gobble}%
\providecommand \bibinfo  [0]{\@secondoftwo}%
\providecommand \bibfield  [0]{\@secondoftwo}%
\providecommand \translation [1]{[#1]}%
\providecommand \BibitemOpen [0]{}%
\providecommand \bibitemStop [0]{}%
\providecommand \bibitemNoStop [0]{.\EOS\space}%
\providecommand \EOS [0]{\spacefactor3000\relax}%
\providecommand \BibitemShut  [1]{\csname bibitem#1\endcsname}%
\let\auto@bib@innerbib\@empty
\bibitem [{\citenamefont {Degen}\ \emph {et~al.}(2017)\citenamefont {Degen}, \citenamefont {Reinhard},\ and\ \citenamefont {Cappellaro}}]{RevModPhys_QSensing}%
  \BibitemOpen
  \bibfield  {author} {\bibinfo {author} {\bibfnamefont {C.~L.}\ \bibnamefont {Degen}}, \bibinfo {author} {\bibfnamefont {F.}~\bibnamefont {Reinhard}},\ and\ \bibinfo {author} {\bibfnamefont {P.}~\bibnamefont {Cappellaro}},\ }\bibfield  {title} {\bibinfo {title} {Quantum sensing},\ }\href {https://doi.org/10.1103/RevModPhys.89.035002} {\bibfield  {journal} {\bibinfo  {journal} {Rev. Mod. Phys.}\ }\textbf {\bibinfo {volume} {89}},\ \bibinfo {pages} {035002} (\bibinfo {year} {2017})}\BibitemShut {NoStop}%
\bibitem [{\citenamefont {Paris}(2009)}]{Paris2009}%
  \BibitemOpen
  \bibfield  {author} {\bibinfo {author} {\bibfnamefont {M.~G.~A.}\ \bibnamefont {Paris}},\ }\bibfield  {title} {\bibinfo {title} {Quantum estimation for quantum technology},\ }\href {https://doi.org/10.1142/S0219749909004839} {\bibfield  {journal} {\bibinfo  {journal} {Int. J. Quantum Inf.}\ }\textbf {\bibinfo {volume} {07}},\ \bibinfo {pages} {125} (\bibinfo {year} {2009})}\BibitemShut {NoStop}%
\bibitem [{\citenamefont {Zanardi}\ \emph {et~al.}(2008)\citenamefont {Zanardi}, \citenamefont {Paris},\ and\ \citenamefont {Campos~Venuti}}]{Zanardi2008}%
  \BibitemOpen
  \bibfield  {author} {\bibinfo {author} {\bibfnamefont {P.}~\bibnamefont {Zanardi}}, \bibinfo {author} {\bibfnamefont {M.~G.~A.}\ \bibnamefont {Paris}},\ and\ \bibinfo {author} {\bibfnamefont {L.}~\bibnamefont {Campos~Venuti}},\ }\bibfield  {title} {\bibinfo {title} {Quantum criticality as a resource for quantum estimation},\ }\href {https://doi.org/10.1103/PhysRevA.78.042105} {\bibfield  {journal} {\bibinfo  {journal} {Phys. Rev. A}\ }\textbf {\bibinfo {volume} {78}},\ \bibinfo {pages} {042105} (\bibinfo {year} {2008})}\BibitemShut {NoStop}%
\bibitem [{\citenamefont {Ivanov}\ and\ \citenamefont {Porras}(2013)}]{ivanov_adiabatic_2013}%
  \BibitemOpen
  \bibfield  {author} {\bibinfo {author} {\bibfnamefont {P.~A.}\ \bibnamefont {Ivanov}}\ and\ \bibinfo {author} {\bibfnamefont {D.}~\bibnamefont {Porras}},\ }\bibfield  {title} {\bibinfo {title} {Adiabatic quantum metrology with strongly correlated quantum optical systems},\ }\href {https://doi.org/10.1103/PhysRevA.88.023803} {\bibfield  {journal} {\bibinfo  {journal} {Phys. Rev. A}\ }\textbf {\bibinfo {volume} {88}},\ \bibinfo {pages} {023803} (\bibinfo {year} {2013})}\BibitemShut {NoStop}%
\bibitem [{\citenamefont {Bina}\ \emph {et~al.}(2016)\citenamefont {Bina}, \citenamefont {Amelio},\ and\ \citenamefont {Paris}}]{Bina2016}%
  \BibitemOpen
  \bibfield  {author} {\bibinfo {author} {\bibfnamefont {M.}~\bibnamefont {Bina}}, \bibinfo {author} {\bibfnamefont {I.}~\bibnamefont {Amelio}},\ and\ \bibinfo {author} {\bibfnamefont {M.~G.~A.}\ \bibnamefont {Paris}},\ }\bibfield  {title} {\bibinfo {title} {Dicke coupling by feasible local measurements at the superradiant quantum phase transition},\ }\href {https://doi.org/10.1103/PhysRevE.93.052118} {\bibfield  {journal} {\bibinfo  {journal} {Phys. Rev. E}\ }\textbf {\bibinfo {volume} {93}},\ \bibinfo {pages} {052118} (\bibinfo {year} {2016})}\BibitemShut {NoStop}%
\bibitem [{\citenamefont {Fern\'andez-Lorenzo}\ and\ \citenamefont {Porras}(2017)}]{Lorenzo2017}%
  \BibitemOpen
  \bibfield  {author} {\bibinfo {author} {\bibfnamefont {S.}~\bibnamefont {Fern\'andez-Lorenzo}}\ and\ \bibinfo {author} {\bibfnamefont {D.}~\bibnamefont {Porras}},\ }\bibfield  {title} {\bibinfo {title} {Quantum sensing close to a dissipative phase transition: Symmetry breaking and criticality as metrological resources},\ }\href {https://doi.org/10.1103/PhysRevA.96.013817} {\bibfield  {journal} {\bibinfo  {journal} {Phys. Rev. A}\ }\textbf {\bibinfo {volume} {96}},\ \bibinfo {pages} {013817} (\bibinfo {year} {2017})}\BibitemShut {NoStop}%
\bibitem [{\citenamefont {Ivanov}(2020)}]{Ivanov2020}%
  \BibitemOpen
  \bibfield  {author} {\bibinfo {author} {\bibfnamefont {P.~A.}\ \bibnamefont {Ivanov}},\ }\bibfield  {title} {\bibinfo {title} {Enhanced two-parameter phase-space-displacement estimation close to a dissipative phase transition},\ }\href {https://doi.org/10.1103/PhysRevA.102.052611} {\bibfield  {journal} {\bibinfo  {journal} {Phys. Rev. A}\ }\textbf {\bibinfo {volume} {102}},\ \bibinfo {pages} {052611} (\bibinfo {year} {2020})}\BibitemShut {NoStop}%
\bibitem [{\citenamefont {Invernizzi}\ \emph {et~al.}(2008)\citenamefont {Invernizzi}, \citenamefont {Korbman}, \citenamefont {Campos~Venuti},\ and\ \citenamefont {Paris}}]{invernizzi2008Optimal}%
  \BibitemOpen
  \bibfield  {author} {\bibinfo {author} {\bibfnamefont {C.}~\bibnamefont {Invernizzi}}, \bibinfo {author} {\bibfnamefont {M.}~\bibnamefont {Korbman}}, \bibinfo {author} {\bibfnamefont {L.}~\bibnamefont {Campos~Venuti}},\ and\ \bibinfo {author} {\bibfnamefont {M.~G.~A.}\ \bibnamefont {Paris}},\ }\bibfield  {title} {\bibinfo {title} {Optimal quantum estimation in spin systems at criticality},\ }\href {https://doi.org/10.1103/PhysRevA.78.042106} {\bibfield  {journal} {\bibinfo  {journal} {Phys. Rev. A}\ }\textbf {\bibinfo {volume} {78}},\ \bibinfo {pages} {042106} (\bibinfo {year} {2008})}\BibitemShut {NoStop}%
\bibitem [{\citenamefont {Mirkhalaf}\ \emph {et~al.}(2020)\citenamefont {Mirkhalaf}, \citenamefont {Witkowska},\ and\ \citenamefont {Lepori}}]{Mirkhalaf2020}%
  \BibitemOpen
  \bibfield  {author} {\bibinfo {author} {\bibfnamefont {S.~S.}\ \bibnamefont {Mirkhalaf}}, \bibinfo {author} {\bibfnamefont {E.}~\bibnamefont {Witkowska}},\ and\ \bibinfo {author} {\bibfnamefont {L.}~\bibnamefont {Lepori}},\ }\bibfield  {title} {\bibinfo {title} {Supersensitive quantum sensor based on criticality in an antiferromagnetic spinor condensate},\ }\href {https://doi.org/10.1103/PhysRevA.101.043609} {\bibfield  {journal} {\bibinfo  {journal} {Phys. Rev. A}\ }\textbf {\bibinfo {volume} {101}},\ \bibinfo {pages} {043609} (\bibinfo {year} {2020})}\BibitemShut {NoStop}%
\bibitem [{\citenamefont {Niezgoda}\ and\ \citenamefont {Chwede\ifmmode~\acute{n}\else \'{n}\fi{}czuk}(2021)}]{Niezgoda2021}%
  \BibitemOpen
  \bibfield  {author} {\bibinfo {author} {\bibfnamefont {A.}~\bibnamefont {Niezgoda}}\ and\ \bibinfo {author} {\bibfnamefont {J.}~\bibnamefont {Chwede\ifmmode~\acute{n}\else \'{n}\fi{}czuk}},\ }\bibfield  {title} {\bibinfo {title} {Many-body nonlocality as a resource for quantum-enhanced metrology},\ }\href {https://doi.org/10.1103/PhysRevLett.126.210506} {\bibfield  {journal} {\bibinfo  {journal} {Phys. Rev. Lett.}\ }\textbf {\bibinfo {volume} {126}},\ \bibinfo {pages} {210506} (\bibinfo {year} {2021})}\BibitemShut {NoStop}%
\bibitem [{\citenamefont {Di~Fresco}\ \emph {et~al.}(2022)\citenamefont {Di~Fresco}, \citenamefont {Spagnolo}, \citenamefont {Valenti},\ and\ \citenamefont {Carollo}}]{DiFresco2022}%
  \BibitemOpen
  \bibfield  {author} {\bibinfo {author} {\bibfnamefont {G.}~\bibnamefont {Di~Fresco}}, \bibinfo {author} {\bibfnamefont {B.}~\bibnamefont {Spagnolo}}, \bibinfo {author} {\bibfnamefont {D.}~\bibnamefont {Valenti}},\ and\ \bibinfo {author} {\bibfnamefont {A.}~\bibnamefont {Carollo}},\ }\bibfield  {title} {\bibinfo {title} {{Multiparameter quantum critical metrology}},\ }\href {https://doi.org/10.21468/SciPostPhys.13.4.077} {\bibfield  {journal} {\bibinfo  {journal} {SciPost Phys.}\ }\textbf {\bibinfo {volume} {13}},\ \bibinfo {pages} {077} (\bibinfo {year} {2022})}\BibitemShut {NoStop}%
\bibitem [{\citenamefont {Di~Fresco}\ \emph {et~al.}(2024)\citenamefont {Di~Fresco}, \citenamefont {Spagnolo}, \citenamefont {Valenti},\ and\ \citenamefont {Carollo}}]{DiFresco2024}%
  \BibitemOpen
  \bibfield  {author} {\bibinfo {author} {\bibfnamefont {G.}~\bibnamefont {Di~Fresco}}, \bibinfo {author} {\bibfnamefont {B.}~\bibnamefont {Spagnolo}}, \bibinfo {author} {\bibfnamefont {D.}~\bibnamefont {Valenti}},\ and\ \bibinfo {author} {\bibfnamefont {A.}~\bibnamefont {Carollo}},\ }\bibfield  {title} {\bibinfo {title} {Metrology and multipartite entanglement in measurement-induced phase transition},\ }\href {https://doi.org/10.22331/q-2024-04-30-1326} {\bibfield  {journal} {\bibinfo  {journal} {Quantum}\ }\textbf {\bibinfo {volume} {8}},\ \bibinfo {pages} {1326} (\bibinfo {year} {2024})}\BibitemShut {NoStop}%
\bibitem [{\citenamefont {Sahoo}\ \emph {et~al.}(2024)\citenamefont {Sahoo}, \citenamefont {Mishra},\ and\ \citenamefont {Rakshit}}]{Sahoo24}%
  \BibitemOpen
  \bibfield  {author} {\bibinfo {author} {\bibfnamefont {A.}~\bibnamefont {Sahoo}}, \bibinfo {author} {\bibfnamefont {U.}~\bibnamefont {Mishra}},\ and\ \bibinfo {author} {\bibfnamefont {D.}~\bibnamefont {Rakshit}},\ }\bibfield  {title} {\bibinfo {title} {Localization-driven quantum sensing},\ }\href {https://doi.org/10.1103/PhysRevA.109.L030601} {\bibfield  {journal} {\bibinfo  {journal} {Phys. Rev. A}\ }\textbf {\bibinfo {volume} {109}},\ \bibinfo {pages} {L030601} (\bibinfo {year} {2024})}\BibitemShut {NoStop}%
\bibitem [{\citenamefont {Tsang}(2013)}]{Tsang2013}%
  \BibitemOpen
  \bibfield  {author} {\bibinfo {author} {\bibfnamefont {M.}~\bibnamefont {Tsang}},\ }\bibfield  {title} {\bibinfo {title} {Quantum transition-edge detectors},\ }\href {https://doi.org/10.1103/PhysRevA.88.021801} {\bibfield  {journal} {\bibinfo  {journal} {Phys. Rev. A}\ }\textbf {\bibinfo {volume} {88}},\ \bibinfo {pages} {021801(R)} (\bibinfo {year} {2013})}\BibitemShut {NoStop}%
\bibitem [{\citenamefont {Macieszczak}\ \emph {et~al.}(2016)\citenamefont {Macieszczak}, \citenamefont {Gu{\c t}{\u a}}, \citenamefont {Lesanovsky},\ and\ \citenamefont {Garrahan}}]{macieszczak_dynamical_2016}%
  \BibitemOpen
  \bibfield  {author} {\bibinfo {author} {\bibfnamefont {K.}~\bibnamefont {Macieszczak}}, \bibinfo {author} {\bibfnamefont {M.}~\bibnamefont {Gu{\c t}{\u a}}}, \bibinfo {author} {\bibfnamefont {I.}~\bibnamefont {Lesanovsky}},\ and\ \bibinfo {author} {\bibfnamefont {J.~P.}\ \bibnamefont {Garrahan}},\ }\bibfield  {title} {\bibinfo {title} {Dynamical phase transitions as a resource for quantum enhanced metrology},\ }\href {https://doi.org/10.1103/PhysRevA.93.022103} {\bibfield  {journal} {\bibinfo  {journal} {Phys. Rev. A}\ }\textbf {\bibinfo {volume} {93}},\ \bibinfo {pages} {022103} (\bibinfo {year} {2016})}\BibitemShut {NoStop}%
\bibitem [{\citenamefont {Cabot}\ \emph {et~al.}(2024)\citenamefont {Cabot}, \citenamefont {Carollo},\ and\ \citenamefont {Lesanovsky}}]{Cabot24}%
  \BibitemOpen
  \bibfield  {author} {\bibinfo {author} {\bibfnamefont {A.}~\bibnamefont {Cabot}}, \bibinfo {author} {\bibfnamefont {F.}~\bibnamefont {Carollo}},\ and\ \bibinfo {author} {\bibfnamefont {I.}~\bibnamefont {Lesanovsky}},\ }\bibfield  {title} {\bibinfo {title} {Continuous sensing and parameter estimation with the boundary time crystal},\ }\href {https://doi.org/10.1103/PhysRevLett.132.050801} {\bibfield  {journal} {\bibinfo  {journal} {Phys. Rev. Lett.}\ }\textbf {\bibinfo {volume} {132}},\ \bibinfo {pages} {050801} (\bibinfo {year} {2024})}\BibitemShut {NoStop}%
\bibitem [{\citenamefont {Zicari}\ \emph {et~al.}(2024)\citenamefont {Zicari}, \citenamefont {Carlesso}, \citenamefont {Trombettoni},\ and\ \citenamefont {Paternostro}}]{zicari2024}%
  \BibitemOpen
  \bibfield  {author} {\bibinfo {author} {\bibfnamefont {G.}~\bibnamefont {Zicari}}, \bibinfo {author} {\bibfnamefont {M.}~\bibnamefont {Carlesso}}, \bibinfo {author} {\bibfnamefont {A.}~\bibnamefont {Trombettoni}},\ and\ \bibinfo {author} {\bibfnamefont {M.}~\bibnamefont {Paternostro}},\ }\bibfield  {title} {\bibinfo {title} {Criticality-amplified quantum probing of a spontaneous collapse model},\ }\href {https://arxiv.org/abs/2407.09304} {\bibfield  {journal} {\bibinfo  {journal} {arXiv preprint arXiv:2407.09304}\ } (\bibinfo {year} {2024})}\BibitemShut {NoStop}%
\bibitem [{\citenamefont {Rams}\ \emph {et~al.}(2018)\citenamefont {Rams}, \citenamefont {Sierant}, \citenamefont {Dutta}, \citenamefont {Horodecki},\ and\ \citenamefont {Zakrzewski}}]{Rams2018}%
  \BibitemOpen
  \bibfield  {author} {\bibinfo {author} {\bibfnamefont {M.~M.}\ \bibnamefont {Rams}}, \bibinfo {author} {\bibfnamefont {P.}~\bibnamefont {Sierant}}, \bibinfo {author} {\bibfnamefont {O.}~\bibnamefont {Dutta}}, \bibinfo {author} {\bibfnamefont {P.}~\bibnamefont {Horodecki}},\ and\ \bibinfo {author} {\bibfnamefont {J.}~\bibnamefont {Zakrzewski}},\ }\bibfield  {title} {\bibinfo {title} {At the {Limits} of {Criticality}-{Based} {Quantum} {Metrology}: {Apparent} {Super}-{Heisenberg} {Scaling} {Revisited}},\ }\href {https://doi.org/10.1103/PhysRevX.8.021022} {\bibfield  {journal} {\bibinfo  {journal} {Phys. Rev. X}\ }\textbf {\bibinfo {volume} {8}},\ \bibinfo {pages} {021022} (\bibinfo {year} {2018})}\BibitemShut {NoStop}%
\bibitem [{\citenamefont {Montenegro}\ \emph {et~al.}(2021)\citenamefont {Montenegro}, \citenamefont {Mishra},\ and\ \citenamefont {Bayat}}]{Montenegro2021}%
  \BibitemOpen
  \bibfield  {author} {\bibinfo {author} {\bibfnamefont {V.}~\bibnamefont {Montenegro}}, \bibinfo {author} {\bibfnamefont {U.}~\bibnamefont {Mishra}},\ and\ \bibinfo {author} {\bibfnamefont {A.}~\bibnamefont {Bayat}},\ }\bibfield  {title} {\bibinfo {title} {Global sensing and its impact for quantum many-body probes with criticality},\ }\href {https://doi.org/10.1103/PhysRevLett.126.200501} {\bibfield  {journal} {\bibinfo  {journal} {Phys. Rev. Lett.}\ }\textbf {\bibinfo {volume} {126}},\ \bibinfo {pages} {200501} (\bibinfo {year} {2021})}\BibitemShut {NoStop}%
\bibitem [{\citenamefont {Salvia}\ \emph {et~al.}(2023)\citenamefont {Salvia}, \citenamefont {Mehboudi},\ and\ \citenamefont {Perarnau-Llobet}}]{Salvia2023}%
  \BibitemOpen
  \bibfield  {author} {\bibinfo {author} {\bibfnamefont {R.}~\bibnamefont {Salvia}}, \bibinfo {author} {\bibfnamefont {M.}~\bibnamefont {Mehboudi}},\ and\ \bibinfo {author} {\bibfnamefont {M.}~\bibnamefont {Perarnau-Llobet}},\ }\bibfield  {title} {\bibinfo {title} {Critical quantum metrology assisted by real-time feedback control},\ }\href {https://doi.org/10.1103/PhysRevLett.130.240803} {\bibfield  {journal} {\bibinfo  {journal} {Phys. Rev. Lett.}\ }\textbf {\bibinfo {volume} {130}},\ \bibinfo {pages} {240803} (\bibinfo {year} {2023})}\BibitemShut {NoStop}%
\bibitem [{\citenamefont {Mukhopadhyay}\ and\ \citenamefont {Bayat}(2024)}]{PhysRevLett.133.120601}%
  \BibitemOpen
  \bibfield  {author} {\bibinfo {author} {\bibfnamefont {C.}~\bibnamefont {Mukhopadhyay}}\ and\ \bibinfo {author} {\bibfnamefont {A.}~\bibnamefont {Bayat}},\ }\bibfield  {title} {\bibinfo {title} {Modular many-body quantum sensors},\ }\href {https://doi.org/10.1103/PhysRevLett.133.120601} {\bibfield  {journal} {\bibinfo  {journal} {Phys. Rev. Lett.}\ }\textbf {\bibinfo {volume} {133}},\ \bibinfo {pages} {120601} (\bibinfo {year} {2024})}\BibitemShut {NoStop}%
\bibitem [{\citenamefont {Ding}\ \emph {et~al.}(2022)\citenamefont {Ding}, \citenamefont {Liu}, \citenamefont {Shi}, \citenamefont {Guo}, \citenamefont {M{\o}lmer},\ and\ \citenamefont {Adams}}]{Ding2022}%
  \BibitemOpen
  \bibfield  {author} {\bibinfo {author} {\bibfnamefont {D.-S.}\ \bibnamefont {Ding}}, \bibinfo {author} {\bibfnamefont {Z.-K.}\ \bibnamefont {Liu}}, \bibinfo {author} {\bibfnamefont {B.-S.}\ \bibnamefont {Shi}}, \bibinfo {author} {\bibfnamefont {G.-C.}\ \bibnamefont {Guo}}, \bibinfo {author} {\bibfnamefont {K.}~\bibnamefont {M{\o}lmer}},\ and\ \bibinfo {author} {\bibfnamefont {C.~S.}\ \bibnamefont {Adams}},\ }\bibfield  {title} {\bibinfo {title} {Enhanced metrology at the critical point of a many-body {R}ydberg atomic system},\ }\href {https://doi.org/10.1038/s41567-022-01777-8} {\bibfield  {journal} {\bibinfo  {journal} {Nat. Phys.}\ }\textbf {\bibinfo {volume} {18}},\ \bibinfo {pages} {1447} (\bibinfo {year} {2022})}\BibitemShut {NoStop}%
\bibitem [{\citenamefont {Liu}\ \emph {et~al.}(2021)\citenamefont {Liu}, \citenamefont {Chen}, \citenamefont {Jiang}, \citenamefont {Yang}, \citenamefont {Wu}, \citenamefont {Li}, \citenamefont {Yuan}, \citenamefont {Peng},\ and\ \citenamefont {Du}}]{Liu2021}%
  \BibitemOpen
  \bibfield  {author} {\bibinfo {author} {\bibfnamefont {R.}~\bibnamefont {Liu}}, \bibinfo {author} {\bibfnamefont {Y.}~\bibnamefont {Chen}}, \bibinfo {author} {\bibfnamefont {M.}~\bibnamefont {Jiang}}, \bibinfo {author} {\bibfnamefont {X.}~\bibnamefont {Yang}}, \bibinfo {author} {\bibfnamefont {Z.}~\bibnamefont {Wu}}, \bibinfo {author} {\bibfnamefont {Y.}~\bibnamefont {Li}}, \bibinfo {author} {\bibfnamefont {H.}~\bibnamefont {Yuan}}, \bibinfo {author} {\bibfnamefont {X.}~\bibnamefont {Peng}},\ and\ \bibinfo {author} {\bibfnamefont {J.}~\bibnamefont {Du}},\ }\bibfield  {title} {\bibinfo {title} {Experimental critical quantum metrology with the {H}eisenberg scaling},\ }\href {https://doi.org/10.1038/s41534-021-00507-x} {\bibfield  {journal} {\bibinfo  {journal} {npj Quantum Inf.}\ }\textbf {\bibinfo {volume} {7}},\ \bibinfo {pages} {170} (\bibinfo {year} {2021})}\BibitemShut {NoStop}%
\bibitem [{\citenamefont {Petrovnin}\ \emph {et~al.}(2024)\citenamefont {Petrovnin}, \citenamefont {Wang}, \citenamefont {Perelshtein}, \citenamefont {Hakonen},\ and\ \citenamefont {Paraoanu}}]{Petrovnin_2024}%
  \BibitemOpen
  \bibfield  {author} {\bibinfo {author} {\bibfnamefont {K.}~\bibnamefont {Petrovnin}}, \bibinfo {author} {\bibfnamefont {J.}~\bibnamefont {Wang}}, \bibinfo {author} {\bibfnamefont {M.}~\bibnamefont {Perelshtein}}, \bibinfo {author} {\bibfnamefont {P.}~\bibnamefont {Hakonen}},\ and\ \bibinfo {author} {\bibfnamefont {G.~S.}\ \bibnamefont {Paraoanu}},\ }\bibfield  {title} {\bibinfo {title} {Microwave photon detection at parametric criticality},\ }\href {https://doi.org/10.1103/PRXQuantum.5.020342} {\bibfield  {journal} {\bibinfo  {journal} {PRX Quantum}\ }\textbf {\bibinfo {volume} {5}},\ \bibinfo {pages} {020342} (\bibinfo {year} {2024})}\BibitemShut {NoStop}%
\bibitem [{\citenamefont {Beaulieu}\ \emph {et~al.}(2024)\citenamefont {Beaulieu}, \citenamefont {Minganti}, \citenamefont {Frasca}, \citenamefont {Scigliuzzo}, \citenamefont {Felicetti}, \citenamefont {Di~Candia},\ and\ \citenamefont {Scarlino}}]{beaulieu2024}%
  \BibitemOpen
  \bibfield  {author} {\bibinfo {author} {\bibfnamefont {G.}~\bibnamefont {Beaulieu}}, \bibinfo {author} {\bibfnamefont {F.}~\bibnamefont {Minganti}}, \bibinfo {author} {\bibfnamefont {S.}~\bibnamefont {Frasca}}, \bibinfo {author} {\bibfnamefont {M.}~\bibnamefont {Scigliuzzo}}, \bibinfo {author} {\bibfnamefont {S.}~\bibnamefont {Felicetti}}, \bibinfo {author} {\bibfnamefont {R.}~\bibnamefont {Di~Candia}},\ and\ \bibinfo {author} {\bibfnamefont {P.}~\bibnamefont {Scarlino}},\ }\bibfield  {title} {\bibinfo {title} {Criticality-enhanced quantum sensing with a parametric superconducting resonator},\ }\href {https://arxiv.org/abs/2409.19968} {\bibfield  {journal} {\bibinfo  {journal} {arXiv preprint arXiv:2409.19968}\ } (\bibinfo {year} {2024})}\BibitemShut {NoStop}%
\bibitem [{\citenamefont {Garbe}\ \emph {et~al.}(2020)\citenamefont {Garbe}, \citenamefont {Bina}, \citenamefont {Keller}, \citenamefont {Paris},\ and\ \citenamefont {Felicetti}}]{Garbe2020}%
  \BibitemOpen
  \bibfield  {author} {\bibinfo {author} {\bibfnamefont {L.}~\bibnamefont {Garbe}}, \bibinfo {author} {\bibfnamefont {M.}~\bibnamefont {Bina}}, \bibinfo {author} {\bibfnamefont {A.}~\bibnamefont {Keller}}, \bibinfo {author} {\bibfnamefont {M.~G.~A.}\ \bibnamefont {Paris}},\ and\ \bibinfo {author} {\bibfnamefont {S.}~\bibnamefont {Felicetti}},\ }\bibfield  {title} {\bibinfo {title} {Critical quantum metrology with a finite-component quantum phase transition},\ }\href {https://doi.org/10.1103/PhysRevLett.124.120504} {\bibfield  {journal} {\bibinfo  {journal} {Phys. Rev. Lett.}\ }\textbf {\bibinfo {volume} {124}},\ \bibinfo {pages} {120504} (\bibinfo {year} {2020})}\BibitemShut {NoStop}%
\bibitem [{\citenamefont {Hwang}\ \emph {et~al.}(2015)\citenamefont {Hwang}, \citenamefont {Puebla},\ and\ \citenamefont {Plenio}}]{hwang_quantum_2015}%
  \BibitemOpen
  \bibfield  {author} {\bibinfo {author} {\bibfnamefont {M.-J.}\ \bibnamefont {Hwang}}, \bibinfo {author} {\bibfnamefont {R.}~\bibnamefont {Puebla}},\ and\ \bibinfo {author} {\bibfnamefont {M.~B.}\ \bibnamefont {Plenio}},\ }\bibfield  {title} {\bibinfo {title} {Quantum {Phase} {Transition} and {Universal} {Dynamics} in the {Rabi} {Model}},\ }\href {https://doi.org/10.1103/PhysRevLett.115.180404} {\bibfield  {journal} {\bibinfo  {journal} {Phys. Rev. Lett.}\ }\textbf {\bibinfo {volume} {115}},\ \bibinfo {pages} {180404} (\bibinfo {year} {2015})}\BibitemShut {NoStop}%
\bibitem [{\citenamefont {Felicetti}\ and\ \citenamefont {Le~Boit{\'e}}(2020)}]{Felicetti2020}%
  \BibitemOpen
  \bibfield  {author} {\bibinfo {author} {\bibfnamefont {S.}~\bibnamefont {Felicetti}}\ and\ \bibinfo {author} {\bibfnamefont {A.}~\bibnamefont {Le~Boit{\'e}}},\ }\bibfield  {title} {\bibinfo {title} {{Universal Spectral Features of Ultrastrongly Coupled Systems}},\ }\href {https://doi.org/10.1103/PhysRevLett.124.040404} {\bibfield  {journal} {\bibinfo  {journal} {Phys. Rev. Lett.}\ }\textbf {\bibinfo {volume} {124}},\ \bibinfo {pages} {040404} (\bibinfo {year} {2020})}\BibitemShut {NoStop}%
\bibitem [{\citenamefont {Ashhab}(2013)}]{Ashhab2013}%
  \BibitemOpen
  \bibfield  {author} {\bibinfo {author} {\bibfnamefont {S.}~\bibnamefont {Ashhab}},\ }\bibfield  {title} {\bibinfo {title} {Superradiance transition in a system with a single qubit and a single oscillator},\ }\href {https://doi.org/10.1103/PhysRevA.87.013826} {\bibfield  {journal} {\bibinfo  {journal} {Phys. Rev. A}\ }\textbf {\bibinfo {volume} {87}},\ \bibinfo {pages} {013826} (\bibinfo {year} {2013})}\BibitemShut {NoStop}%
\bibitem [{\citenamefont {Puebla}\ \emph {et~al.}(2017)\citenamefont {Puebla}, \citenamefont {Hwang}, \citenamefont {Casanova},\ and\ \citenamefont {Plenio}}]{Puebla2017}%
  \BibitemOpen
  \bibfield  {author} {\bibinfo {author} {\bibfnamefont {R.}~\bibnamefont {Puebla}}, \bibinfo {author} {\bibfnamefont {M.-J.}\ \bibnamefont {Hwang}}, \bibinfo {author} {\bibfnamefont {J.}~\bibnamefont {Casanova}},\ and\ \bibinfo {author} {\bibfnamefont {M.~B.}\ \bibnamefont {Plenio}},\ }\bibfield  {title} {\bibinfo {title} {Probing the dynamics of a superradiant quantum phase transition with a single trapped ion},\ }\href {https://doi.org/10.1103/PhysRevLett.118.073001} {\bibfield  {journal} {\bibinfo  {journal} {Phys. Rev. Lett.}\ }\textbf {\bibinfo {volume} {118}},\ \bibinfo {pages} {073001} (\bibinfo {year} {2017})}\BibitemShut {NoStop}%
\bibitem [{\citenamefont {Peng}\ \emph {et~al.}(2019)\citenamefont {Peng}, \citenamefont {Rico}, \citenamefont {Zhong}, \citenamefont {Solano},\ and\ \citenamefont {Egusquiza}}]{Peng2019}%
  \BibitemOpen
  \bibfield  {author} {\bibinfo {author} {\bibfnamefont {J.}~\bibnamefont {Peng}}, \bibinfo {author} {\bibfnamefont {E.}~\bibnamefont {Rico}}, \bibinfo {author} {\bibfnamefont {J.}~\bibnamefont {Zhong}}, \bibinfo {author} {\bibfnamefont {E.}~\bibnamefont {Solano}},\ and\ \bibinfo {author} {\bibfnamefont {I.~L.}\ \bibnamefont {Egusquiza}},\ }\bibfield  {title} {\bibinfo {title} {Unified superradiant phase transitions},\ }\href {https://doi.org/10.1103/PhysRevA.100.063820} {\bibfield  {journal} {\bibinfo  {journal} {Phys. Rev. A}\ }\textbf {\bibinfo {volume} {100}},\ \bibinfo {pages} {063820} (\bibinfo {year} {2019})}\BibitemShut {NoStop}%
\bibitem [{\citenamefont {Zhu}\ \emph {et~al.}(2020)\citenamefont {Zhu}, \citenamefont {Xu}, \citenamefont {Zhang},\ and\ \citenamefont {Liu}}]{Zhu2020}%
  \BibitemOpen
  \bibfield  {author} {\bibinfo {author} {\bibfnamefont {H.-J.}\ \bibnamefont {Zhu}}, \bibinfo {author} {\bibfnamefont {K.}~\bibnamefont {Xu}}, \bibinfo {author} {\bibfnamefont {G.-F.}\ \bibnamefont {Zhang}},\ and\ \bibinfo {author} {\bibfnamefont {W.-M.}\ \bibnamefont {Liu}},\ }\bibfield  {title} {\bibinfo {title} {{Finite-Component Multicriticality at the Superradiant Quantum Phase Transition}},\ }\href {https://doi.org/10.1103/PhysRevLett.125.050402} {\bibfield  {journal} {\bibinfo  {journal} {Phys. Rev. Lett.}\ }\textbf {\bibinfo {volume} {125}},\ \bibinfo {pages} {050402} (\bibinfo {year} {2020})}\BibitemShut {NoStop}%
\bibitem [{\citenamefont {Zhang}\ \emph {et~al.}(2024)\citenamefont {Zhang}, \citenamefont {L\"{u}}, \citenamefont {Chen}, \citenamefont {Yu}, \citenamefont {Wu}, \citenamefont {Yang},\ and\ \citenamefont {Zheng}}]{Zhang24}%
  \BibitemOpen
  \bibfield  {author} {\bibinfo {author} {\bibfnamefont {H.-L.}\ \bibnamefont {Zhang}}, \bibinfo {author} {\bibfnamefont {J.-H.}\ \bibnamefont {L\"{u}}}, \bibinfo {author} {\bibfnamefont {K.}~\bibnamefont {Chen}}, \bibinfo {author} {\bibfnamefont {X.-J.}\ \bibnamefont {Yu}}, \bibinfo {author} {\bibfnamefont {F.}~\bibnamefont {Wu}}, \bibinfo {author} {\bibfnamefont {Z.-B.}\ \bibnamefont {Yang}},\ and\ \bibinfo {author} {\bibfnamefont {S.-B.}\ \bibnamefont {Zheng}},\ }\bibfield  {title} {\bibinfo {title} {Critical quantum geometric tensors of parametrically-driven nonlinear resonators},\ }\href {https://doi.org/10.1364/OE.517716} {\bibfield  {journal} {\bibinfo  {journal} {Opt. Express}\ }\textbf {\bibinfo {volume} {32}},\ \bibinfo {pages} {22566} (\bibinfo {year} {2024})}\BibitemShut {NoStop}%
\bibitem [{\citenamefont {Bartolo}\ \emph {et~al.}(2016)\citenamefont {Bartolo}, \citenamefont {Minganti}, \citenamefont {Casteels},\ and\ \citenamefont {Ciuti}}]{Bartolo2016}%
  \BibitemOpen
  \bibfield  {author} {\bibinfo {author} {\bibfnamefont {N.}~\bibnamefont {Bartolo}}, \bibinfo {author} {\bibfnamefont {F.}~\bibnamefont {Minganti}}, \bibinfo {author} {\bibfnamefont {W.}~\bibnamefont {Casteels}},\ and\ \bibinfo {author} {\bibfnamefont {C.}~\bibnamefont {Ciuti}},\ }\bibfield  {title} {\bibinfo {title} {Exact steady state of a {K}err resonator with one- and two-photon driving and dissipation: {C}ontrollable {W}igner-function multimodality and dissipative phase transitions},\ }\href {https://doi.org/10.1103/PhysRevA.94.033841} {\bibfield  {journal} {\bibinfo  {journal} {Phys. Rev. A}\ }\textbf {\bibinfo {volume} {94}},\ \bibinfo {pages} {033841} (\bibinfo {year} {2016})}\BibitemShut {NoStop}%
\bibitem [{\citenamefont {Minganti}\ \emph {et~al.}(2023{\natexlab{a}})\citenamefont {Minganti}, \citenamefont {Garbe}, \citenamefont {Le~Boit{\'e}},\ and\ \citenamefont {Felicetti}}]{Minganti2023a}%
  \BibitemOpen
  \bibfield  {author} {\bibinfo {author} {\bibfnamefont {F.}~\bibnamefont {Minganti}}, \bibinfo {author} {\bibfnamefont {L.}~\bibnamefont {Garbe}}, \bibinfo {author} {\bibfnamefont {A.}~\bibnamefont {Le~Boit{\'e}}},\ and\ \bibinfo {author} {\bibfnamefont {S.}~\bibnamefont {Felicetti}},\ }\bibfield  {title} {\bibinfo {title} {Non-{G}aussian superradiant transition via three-body ultrastrong coupling},\ }\href {https://doi.org/10.1103/PhysRevA.107.013715} {\bibfield  {journal} {\bibinfo  {journal} {Phys. Rev. A}\ }\textbf {\bibinfo {volume} {107}},\ \bibinfo {pages} {013715} (\bibinfo {year} {2023}{\natexlab{a}})}\BibitemShut {NoStop}%
\bibitem [{\citenamefont {Minganti}\ \emph {et~al.}(2023{\natexlab{b}})\citenamefont {Minganti}, \citenamefont {Savona},\ and\ \citenamefont {Biella}}]{Minganti2023b}%
  \BibitemOpen
  \bibfield  {author} {\bibinfo {author} {\bibfnamefont {F.}~\bibnamefont {Minganti}}, \bibinfo {author} {\bibfnamefont {V.}~\bibnamefont {Savona}},\ and\ \bibinfo {author} {\bibfnamefont {A.}~\bibnamefont {Biella}},\ }\bibfield  {title} {\bibinfo {title} {Dissipative phase transitions in $ n $-photon driven quantum nonlinear resonators},\ }\href {https://doi.org/10.22331/q-2023-11-07-1170} {\bibfield  {journal} {\bibinfo  {journal} {Quantum}\ }\textbf {\bibinfo {volume} {7}},\ \bibinfo {pages} {1170} (\bibinfo {year} {2023}{\natexlab{b}})}\BibitemShut {NoStop}%
\bibitem [{\citenamefont {Chu}\ \emph {et~al.}(2021)\citenamefont {Chu}, \citenamefont {Zhang}, \citenamefont {Yu},\ and\ \citenamefont {Cai}}]{Chu2021}%
  \BibitemOpen
  \bibfield  {author} {\bibinfo {author} {\bibfnamefont {Y.}~\bibnamefont {Chu}}, \bibinfo {author} {\bibfnamefont {S.}~\bibnamefont {Zhang}}, \bibinfo {author} {\bibfnamefont {B.}~\bibnamefont {Yu}},\ and\ \bibinfo {author} {\bibfnamefont {J.}~\bibnamefont {Cai}},\ }\bibfield  {title} {\bibinfo {title} {Dynamic framework for criticality-enhanced quantum sensing},\ }\href {https://doi.org/10.1103/PhysRevLett.126.010502} {\bibfield  {journal} {\bibinfo  {journal} {Phys. Rev. Lett.}\ }\textbf {\bibinfo {volume} {126}},\ \bibinfo {pages} {010502} (\bibinfo {year} {2021})}\BibitemShut {NoStop}%
\bibitem [{\citenamefont {Garbe}\ \emph {et~al.}(2022{\natexlab{a}})\citenamefont {Garbe}, \citenamefont {Abah}, \citenamefont {Felicetti},\ and\ \citenamefont {Puebla}}]{Garbe2022}%
  \BibitemOpen
  \bibfield  {author} {\bibinfo {author} {\bibfnamefont {L.}~\bibnamefont {Garbe}}, \bibinfo {author} {\bibfnamefont {O.}~\bibnamefont {Abah}}, \bibinfo {author} {\bibfnamefont {S.}~\bibnamefont {Felicetti}},\ and\ \bibinfo {author} {\bibfnamefont {P.}~\bibnamefont {Puebla}},\ }\bibfield  {title} {\bibinfo {title} {Critical quantum metrology with fully-connected models: from {H}eisenberg to {K}ibble–{Z}urek scaling},\ }\href {https://doi.org/10.1088/2058-9565/ac6ca5} {\bibfield  {journal} {\bibinfo  {journal} {Quantum Sci. Technol.}\ }\textbf {\bibinfo {volume} {7}},\ \bibinfo {pages} {035010} (\bibinfo {year} {2022}{\natexlab{a}})}\BibitemShut {NoStop}%
\bibitem [{\citenamefont {Gietka}\ \emph {et~al.}(2022)\citenamefont {Gietka}, \citenamefont {Ruks},\ and\ \citenamefont {Busch}}]{Gietka2022}%
  \BibitemOpen
  \bibfield  {author} {\bibinfo {author} {\bibfnamefont {K.}~\bibnamefont {Gietka}}, \bibinfo {author} {\bibfnamefont {L.}~\bibnamefont {Ruks}},\ and\ \bibinfo {author} {\bibfnamefont {T.}~\bibnamefont {Busch}},\ }\bibfield  {title} {\bibinfo {title} {Understanding and improving critical metrology. quenching superradiant light-matter systems beyond the critical point},\ }\href {https://quantum-journal.org/papers/q-2022-04-27-700/} {\bibfield  {journal} {\bibinfo  {journal} {Quantum}\ }\textbf {\bibinfo {volume} {6}},\ \bibinfo {pages} {700} (\bibinfo {year} {2022})}\BibitemShut {NoStop}%
\bibitem [{\citenamefont {Gietka}(2022)}]{Gietka2022a}%
  \BibitemOpen
  \bibfield  {author} {\bibinfo {author} {\bibfnamefont {K.}~\bibnamefont {Gietka}},\ }\bibfield  {title} {\bibinfo {title} {Squeezing by critical speeding up: Applications in quantum metrology},\ }\href {https://doi.org/10.1103/PhysRevA.105.042620} {\bibfield  {journal} {\bibinfo  {journal} {Phys. Rev. A}\ }\textbf {\bibinfo {volume} {105}},\ \bibinfo {pages} {042620} (\bibinfo {year} {2022})}\BibitemShut {NoStop}%
\bibitem [{\citenamefont {Garbe}\ \emph {et~al.}(2022{\natexlab{b}})\citenamefont {Garbe}, \citenamefont {Abah}, \citenamefont {Felicetti},\ and\ \citenamefont {Puebla}}]{Garbe2022a}%
  \BibitemOpen
  \bibfield  {author} {\bibinfo {author} {\bibfnamefont {L.}~\bibnamefont {Garbe}}, \bibinfo {author} {\bibfnamefont {O.}~\bibnamefont {Abah}}, \bibinfo {author} {\bibfnamefont {S.}~\bibnamefont {Felicetti}},\ and\ \bibinfo {author} {\bibfnamefont {R.}~\bibnamefont {Puebla}},\ }\bibfield  {title} {\bibinfo {title} {Exponential time-scaling of estimation precision by reaching a quantum critical point},\ }\href {https://doi.org/10.1103/PhysRevResearch.4.043061} {\bibfield  {journal} {\bibinfo  {journal} {Phys. Rev. Res.}\ }\textbf {\bibinfo {volume} {4}},\ \bibinfo {pages} {043061} (\bibinfo {year} {2022}{\natexlab{b}})}\BibitemShut {NoStop}%
\bibitem [{\citenamefont {Ilias}\ \emph {et~al.}(2022)\citenamefont {Ilias}, \citenamefont {Yang}, \citenamefont {Huelga},\ and\ \citenamefont {Plenio}}]{Ilias2022}%
  \BibitemOpen
  \bibfield  {author} {\bibinfo {author} {\bibfnamefont {T.}~\bibnamefont {Ilias}}, \bibinfo {author} {\bibfnamefont {D.}~\bibnamefont {Yang}}, \bibinfo {author} {\bibfnamefont {S.~F.}\ \bibnamefont {Huelga}},\ and\ \bibinfo {author} {\bibfnamefont {M.~B.}\ \bibnamefont {Plenio}},\ }\bibfield  {title} {\bibinfo {title} {Criticality-enhanced quantum sensing via continuous measurement},\ }\href {https://doi.org/10.1103/PRXQuantum.3.010354} {\bibfield  {journal} {\bibinfo  {journal} {PRX Quantum}\ }\textbf {\bibinfo {volume} {3}},\ \bibinfo {pages} {010354} (\bibinfo {year} {2022})}\BibitemShut {NoStop}%
\bibitem [{\citenamefont {Yang}\ \emph {et~al.}(2023)\citenamefont {Yang}, \citenamefont {Huelga},\ and\ \citenamefont {Plenio}}]{Yang2022}%
  \BibitemOpen
  \bibfield  {author} {\bibinfo {author} {\bibfnamefont {D.}~\bibnamefont {Yang}}, \bibinfo {author} {\bibfnamefont {S.~F.}\ \bibnamefont {Huelga}},\ and\ \bibinfo {author} {\bibfnamefont {M.~B.}\ \bibnamefont {Plenio}},\ }\bibfield  {title} {\bibinfo {title} {Efficient information retrieval for sensing via continuous measurement},\ }\href {https://doi.org/10.1103/PhysRevX.13.031012} {\bibfield  {journal} {\bibinfo  {journal} {Phys. Rev. X}\ }\textbf {\bibinfo {volume} {13}},\ \bibinfo {pages} {031012} (\bibinfo {year} {2023})}\BibitemShut {NoStop}%
\bibitem [{\citenamefont {Alushi}\ \emph {et~al.}(2024)\citenamefont {Alushi}, \citenamefont {G\'orecki}, \citenamefont {Felicetti},\ and\ \citenamefont {Di~Candia}}]{alushi2024optimality}%
  \BibitemOpen
  \bibfield  {author} {\bibinfo {author} {\bibfnamefont {U.}~\bibnamefont {Alushi}}, \bibinfo {author} {\bibfnamefont {W.}~\bibnamefont {G\'orecki}}, \bibinfo {author} {\bibfnamefont {S.}~\bibnamefont {Felicetti}},\ and\ \bibinfo {author} {\bibfnamefont {R.}~\bibnamefont {Di~Candia}},\ }\bibfield  {title} {\bibinfo {title} {Optimality and noise resilience of critical quantum sensing},\ }\href {https://doi.org/10.1103/PhysRevLett.133.040801} {\bibfield  {journal} {\bibinfo  {journal} {Phys. Rev. Lett.}\ }\textbf {\bibinfo {volume} {133}},\ \bibinfo {pages} {040801} (\bibinfo {year} {2024})}\BibitemShut {NoStop}%
\bibitem [{\citenamefont {Cai}\ \emph {et~al.}(2021)\citenamefont {Cai}, \citenamefont {Liu}, \citenamefont {Zhao}, \citenamefont {Wu}, \citenamefont {Mei}, \citenamefont {Jiang}, \citenamefont {He}, \citenamefont {Zhang}, \citenamefont {Zhou},\ and\ \citenamefont {Duan}}]{cai2021observation}%
  \BibitemOpen
  \bibfield  {author} {\bibinfo {author} {\bibfnamefont {M.-L.}\ \bibnamefont {Cai}}, \bibinfo {author} {\bibfnamefont {Z.-D.}\ \bibnamefont {Liu}}, \bibinfo {author} {\bibfnamefont {W.-D.}\ \bibnamefont {Zhao}}, \bibinfo {author} {\bibfnamefont {Y.-K.}\ \bibnamefont {Wu}}, \bibinfo {author} {\bibfnamefont {Q.-X.}\ \bibnamefont {Mei}}, \bibinfo {author} {\bibfnamefont {Y.}~\bibnamefont {Jiang}}, \bibinfo {author} {\bibfnamefont {L.}~\bibnamefont {He}}, \bibinfo {author} {\bibfnamefont {X.}~\bibnamefont {Zhang}}, \bibinfo {author} {\bibfnamefont {Z.-C.}\ \bibnamefont {Zhou}},\ and\ \bibinfo {author} {\bibfnamefont {L.-M.}\ \bibnamefont {Duan}},\ }\bibfield  {title} {\bibinfo {title} {Observation of a quantum phase transition in the quantum {R}abi model with a single trapped ion},\ }\href {https://doi.org/https://doi.org/10.1038/s41467-021-21425-8} {\bibfield  {journal} {\bibinfo  {journal} {Nat. Commun.}\ }\textbf {\bibinfo {volume} {12}},\ \bibinfo {pages} {1126} (\bibinfo {year} {2021})}\BibitemShut
  {NoStop}%
\bibitem [{\citenamefont {Delteil}\ \emph {et~al.}(2019)\citenamefont {Delteil}, \citenamefont {Fink}, \citenamefont {Schade}, \citenamefont {H{\"o}fling}, \citenamefont {Schneider},\ and\ \citenamefont {Imamo\v{g}lu}}]{DelteilNatMat19}%
  \BibitemOpen
  \bibfield  {author} {\bibinfo {author} {\bibfnamefont {A.}~\bibnamefont {Delteil}}, \bibinfo {author} {\bibfnamefont {T.}~\bibnamefont {Fink}}, \bibinfo {author} {\bibfnamefont {A.}~\bibnamefont {Schade}}, \bibinfo {author} {\bibfnamefont {S.}~\bibnamefont {H{\"o}fling}}, \bibinfo {author} {\bibfnamefont {C.}~\bibnamefont {Schneider}},\ and\ \bibinfo {author} {\bibfnamefont {A.}~\bibnamefont {Imamo\v{g}lu}},\ }\bibfield  {title} {\bibinfo {title} {Towards polariton blockade of confined exciton-polaritons},\ }\href {https://doi.org/10.1038/s41563-019-0282-y} {\bibfield  {journal} {\bibinfo  {journal} {Nat. Mat.}\ }\textbf {\bibinfo {volume} {18}},\ \bibinfo {pages} {219} (\bibinfo {year} {2019})}\BibitemShut {NoStop}%
\bibitem [{\citenamefont {Li}\ \emph {et~al.}(2022)\citenamefont {Li}, \citenamefont {Claude}, \citenamefont {Boulier}, \citenamefont {Giacobino}, \citenamefont {Glorieux}, \citenamefont {Bramati},\ and\ \citenamefont {Ciuti}}]{Zejian2022}%
  \BibitemOpen
  \bibfield  {author} {\bibinfo {author} {\bibfnamefont {Z.}~\bibnamefont {Li}}, \bibinfo {author} {\bibfnamefont {F.}~\bibnamefont {Claude}}, \bibinfo {author} {\bibfnamefont {T.}~\bibnamefont {Boulier}}, \bibinfo {author} {\bibfnamefont {E.}~\bibnamefont {Giacobino}}, \bibinfo {author} {\bibfnamefont {Q.}~\bibnamefont {Glorieux}}, \bibinfo {author} {\bibfnamefont {A.}~\bibnamefont {Bramati}},\ and\ \bibinfo {author} {\bibfnamefont {C.}~\bibnamefont {Ciuti}},\ }\bibfield  {title} {\bibinfo {title} {Dissipative phase transition with driving-controlled spatial dimension and diffusive boundary conditions},\ }\href {https://doi.org/10.1103/PhysRevLett.128.093601} {\bibfield  {journal} {\bibinfo  {journal} {Phys. Rev. Lett.}\ }\textbf {\bibinfo {volume} {128}},\ \bibinfo {pages} {093601} (\bibinfo {year} {2022})}\BibitemShut {NoStop}%
\bibitem [{\citenamefont {Fink}\ \emph {et~al.}(2017)\citenamefont {Fink}, \citenamefont {Dombi}, \citenamefont {Vukics}, \citenamefont {Wallraff},\ and\ \citenamefont {Domokos}}]{Fink2017}%
  \BibitemOpen
  \bibfield  {author} {\bibinfo {author} {\bibfnamefont {J.~M.}\ \bibnamefont {Fink}}, \bibinfo {author} {\bibfnamefont {A.}~\bibnamefont {Dombi}}, \bibinfo {author} {\bibfnamefont {A.}~\bibnamefont {Vukics}}, \bibinfo {author} {\bibfnamefont {A.}~\bibnamefont {Wallraff}},\ and\ \bibinfo {author} {\bibfnamefont {P.}~\bibnamefont {Domokos}},\ }\bibfield  {title} {\bibinfo {title} {Observation of the photon-blockade breakdown phase transition},\ }\href {https://doi.org/10.1103/PhysRevX.7.011012} {\bibfield  {journal} {\bibinfo  {journal} {Phys. Rev. X}\ }\textbf {\bibinfo {volume} {7}},\ \bibinfo {pages} {011012} (\bibinfo {year} {2017})}\BibitemShut {NoStop}%
\bibitem [{\citenamefont {Brookes}\ \emph {et~al.}(2021)\citenamefont {Brookes}, \citenamefont {Tancredi}, \citenamefont {Patterson}, \citenamefont {Rahamim}, \citenamefont {Esposito}, \citenamefont {Mavrogordatos}, \citenamefont {Leek}, \citenamefont {Ginossar},\ and\ \citenamefont {Szymanska}}]{Brooks2021}%
  \BibitemOpen
  \bibfield  {author} {\bibinfo {author} {\bibfnamefont {P.}~\bibnamefont {Brookes}}, \bibinfo {author} {\bibfnamefont {G.}~\bibnamefont {Tancredi}}, \bibinfo {author} {\bibfnamefont {A.~D.}\ \bibnamefont {Patterson}}, \bibinfo {author} {\bibfnamefont {J.}~\bibnamefont {Rahamim}}, \bibinfo {author} {\bibfnamefont {M.}~\bibnamefont {Esposito}}, \bibinfo {author} {\bibfnamefont {T.~K.}\ \bibnamefont {Mavrogordatos}}, \bibinfo {author} {\bibfnamefont {P.~J.}\ \bibnamefont {Leek}}, \bibinfo {author} {\bibfnamefont {E.}~\bibnamefont {Ginossar}},\ and\ \bibinfo {author} {\bibfnamefont {M.~H.}\ \bibnamefont {Szymanska}},\ }\bibfield  {title} {\bibinfo {title} {Critical slowing down in circuit quantum electrodynamics},\ }\href {https://doi.org/10.1126/sciadv.abe9492} {\bibfield  {journal} {\bibinfo  {journal} {Sci. Adv.}\ }\textbf {\bibinfo {volume} {7}},\ \bibinfo {pages} {eabe9492} (\bibinfo {year} {2021})}\BibitemShut {NoStop}%
\bibitem [{\citenamefont {Beaulieu}\ \emph {et~al.}(2023)\citenamefont {Beaulieu}, \citenamefont {Minganti}, \citenamefont {Frasca}, \citenamefont {Savona}, \citenamefont {Felicetti}, \citenamefont {Di~Candia},\ and\ \citenamefont {Scarlino}}]{beaulieu2023observation}%
  \BibitemOpen
  \bibfield  {author} {\bibinfo {author} {\bibfnamefont {G.}~\bibnamefont {Beaulieu}}, \bibinfo {author} {\bibfnamefont {F.}~\bibnamefont {Minganti}}, \bibinfo {author} {\bibfnamefont {S.}~\bibnamefont {Frasca}}, \bibinfo {author} {\bibfnamefont {V.}~\bibnamefont {Savona}}, \bibinfo {author} {\bibfnamefont {S.}~\bibnamefont {Felicetti}}, \bibinfo {author} {\bibfnamefont {R.}~\bibnamefont {Di~Candia}},\ and\ \bibinfo {author} {\bibfnamefont {P.}~\bibnamefont {Scarlino}},\ }\bibfield  {title} {\bibinfo {title} {Observation of first-and second-order dissipative phase transitions in a two-photon driven {K}err resonator},\ }\href {https://arxiv.org/abs/2310.13636} {\bibfield  {journal} {\bibinfo  {journal} {arXiv preprint arXiv:2310.13636}\ } (\bibinfo {year} {2023})}\BibitemShut {NoStop}%
\bibitem [{\citenamefont {Chen}\ \emph {et~al.}(2023)\citenamefont {Chen}, \citenamefont {Fischer}, \citenamefont {Nojiri}, \citenamefont {Renger}, \citenamefont {Xie}, \citenamefont {Partanen}, \citenamefont {Pogorzalek}, \citenamefont {Fedorov}, \citenamefont {Marx}, \citenamefont {Deppe} \emph {et~al.}}]{chen2023quantum}%
  \BibitemOpen
  \bibfield  {author} {\bibinfo {author} {\bibfnamefont {Q.-M.}\ \bibnamefont {Chen}}, \bibinfo {author} {\bibfnamefont {M.}~\bibnamefont {Fischer}}, \bibinfo {author} {\bibfnamefont {Y.}~\bibnamefont {Nojiri}}, \bibinfo {author} {\bibfnamefont {M.}~\bibnamefont {Renger}}, \bibinfo {author} {\bibfnamefont {E.}~\bibnamefont {Xie}}, \bibinfo {author} {\bibfnamefont {M.}~\bibnamefont {Partanen}}, \bibinfo {author} {\bibfnamefont {S.}~\bibnamefont {Pogorzalek}}, \bibinfo {author} {\bibfnamefont {K.~G.}\ \bibnamefont {Fedorov}}, \bibinfo {author} {\bibfnamefont {A.}~\bibnamefont {Marx}}, \bibinfo {author} {\bibfnamefont {F.}~\bibnamefont {Deppe}}, \emph {et~al.},\ }\bibfield  {title} {\bibinfo {title} {Quantum behavior of the {D}uffing oscillator at the dissipative phase transition},\ }\href {https://www.nature.com/articles/s41467-023-38217-x#citeas} {\bibfield  {journal} {\bibinfo  {journal} {Nat. Commun.}\ }\textbf {\bibinfo {volume} {14}},\ \bibinfo {pages} {2896} (\bibinfo {year} {2023})}\BibitemShut {NoStop}%
\bibitem [{\citenamefont {Sett}\ \emph {et~al.}(2024)\citenamefont {Sett}, \citenamefont {Hassani}, \citenamefont {Phan}, \citenamefont {Barzanjeh}, \citenamefont {Vukics},\ and\ \citenamefont {Fink}}]{Sett24}%
  \BibitemOpen
  \bibfield  {author} {\bibinfo {author} {\bibfnamefont {R.}~\bibnamefont {Sett}}, \bibinfo {author} {\bibfnamefont {F.}~\bibnamefont {Hassani}}, \bibinfo {author} {\bibfnamefont {D.}~\bibnamefont {Phan}}, \bibinfo {author} {\bibfnamefont {S.}~\bibnamefont {Barzanjeh}}, \bibinfo {author} {\bibfnamefont {A.}~\bibnamefont {Vukics}},\ and\ \bibinfo {author} {\bibfnamefont {J.~M.}\ \bibnamefont {Fink}},\ }\bibfield  {title} {\bibinfo {title} {Emergent macroscopic bistability induced by a single superconducting qubit},\ }\href {https://doi.org/10.1103/PRXQuantum.5.010327} {\bibfield  {journal} {\bibinfo  {journal} {PRX Quantum}\ }\textbf {\bibinfo {volume} {5}},\ \bibinfo {pages} {010327} (\bibinfo {year} {2024})}\BibitemShut {NoStop}%
\bibitem [{\citenamefont {Jouanny}\ \emph {et~al.}(2024)\citenamefont {Jouanny}, \citenamefont {Frasca}, \citenamefont {Weibel}, \citenamefont {Peyruchat}, \citenamefont {Scigliuzzo}, \citenamefont {Oppliger}, \citenamefont {De~Palma}, \citenamefont {Sbroggio}, \citenamefont {Beaulieu}, \citenamefont {Zilberberg} \emph {et~al.}}]{jouanny2024}%
  \BibitemOpen
  \bibfield  {author} {\bibinfo {author} {\bibfnamefont {V.}~\bibnamefont {Jouanny}}, \bibinfo {author} {\bibfnamefont {S.}~\bibnamefont {Frasca}}, \bibinfo {author} {\bibfnamefont {V.~J.}\ \bibnamefont {Weibel}}, \bibinfo {author} {\bibfnamefont {L.}~\bibnamefont {Peyruchat}}, \bibinfo {author} {\bibfnamefont {M.}~\bibnamefont {Scigliuzzo}}, \bibinfo {author} {\bibfnamefont {F.}~\bibnamefont {Oppliger}}, \bibinfo {author} {\bibfnamefont {F.}~\bibnamefont {De~Palma}}, \bibinfo {author} {\bibfnamefont {D.}~\bibnamefont {Sbroggio}}, \bibinfo {author} {\bibfnamefont {G.}~\bibnamefont {Beaulieu}}, \bibinfo {author} {\bibfnamefont {O.}~\bibnamefont {Zilberberg}}, \emph {et~al.},\ }\bibfield  {title} {\bibinfo {title} {Band engineering and study of disorder using topology in compact high kinetic inductance cavity arrays},\ }\href {https://arxiv.org/abs/2403.18150} {\bibfield  {journal} {\bibinfo  {journal} {arXiv preprint arXiv:2403.18150}\ } (\bibinfo {year} {2024})}\BibitemShut {NoStop}%
\bibitem [{\citenamefont {Heugel}\ \emph {et~al.}(2019)\citenamefont {Heugel}, \citenamefont {Biondi}, \citenamefont {Zilberberg},\ and\ \citenamefont {Chitra}}]{heugel2020_quantum}%
  \BibitemOpen
  \bibfield  {author} {\bibinfo {author} {\bibfnamefont {T.~L.}\ \bibnamefont {Heugel}}, \bibinfo {author} {\bibfnamefont {M.}~\bibnamefont {Biondi}}, \bibinfo {author} {\bibfnamefont {O.}~\bibnamefont {Zilberberg}},\ and\ \bibinfo {author} {\bibfnamefont {R.}~\bibnamefont {Chitra}},\ }\bibfield  {title} {\bibinfo {title} {Quantum transducer using a parametric driven-dissipative phase transition},\ }\href {https://doi.org/10.1103/PhysRevLett.123.173601} {\bibfield  {journal} {\bibinfo  {journal} {Phys. Rev. Lett.}\ }\textbf {\bibinfo {volume} {123}},\ \bibinfo {pages} {173601} (\bibinfo {year} {2019})}\BibitemShut {NoStop}%
\bibitem [{\citenamefont {Di~Candia}\ \emph {et~al.}(2023)\citenamefont {Di~Candia}, \citenamefont {Minganti}, \citenamefont {Petrovnin}, \citenamefont {Paraoanu},\ and\ \citenamefont {Felicetti}}]{DiCandia2023}%
  \BibitemOpen
  \bibfield  {author} {\bibinfo {author} {\bibfnamefont {R.}~\bibnamefont {Di~Candia}}, \bibinfo {author} {\bibfnamefont {F.}~\bibnamefont {Minganti}}, \bibinfo {author} {\bibfnamefont {K.~V.}\ \bibnamefont {Petrovnin}}, \bibinfo {author} {\bibfnamefont {G.~S.}\ \bibnamefont {Paraoanu}},\ and\ \bibinfo {author} {\bibfnamefont {S.}~\bibnamefont {Felicetti}},\ }\bibfield  {title} {\bibinfo {title} {Critical parametric quantum sensing},\ }\href {https://doi.org/10.1038/s41534-023-00690-z} {\bibfield  {journal} {\bibinfo  {journal} {npj Quantum Inf.}\ }\textbf {\bibinfo {volume} {9}},\ \bibinfo {pages} {23} (\bibinfo {year} {2023})}\BibitemShut {NoStop}%
\bibitem [{\citenamefont {Rinaldi}\ \emph {et~al.}(2021)\citenamefont {Rinaldi}, \citenamefont {Di~Candia}, \citenamefont {Felicetti},\ and\ \citenamefont {Minganti}}]{Rinaldi2021}%
  \BibitemOpen
  \bibfield  {author} {\bibinfo {author} {\bibfnamefont {E.}~\bibnamefont {Rinaldi}}, \bibinfo {author} {\bibfnamefont {R.}~\bibnamefont {Di~Candia}}, \bibinfo {author} {\bibfnamefont {S.}~\bibnamefont {Felicetti}},\ and\ \bibinfo {author} {\bibfnamefont {F.}~\bibnamefont {Minganti}},\ }\bibfield  {title} {\bibinfo {title} {Dispersive qubit readout with machine learning},\ }\href {https://doi.org/10.48550/arXiv.2112.05332} {\bibfield  {journal} {\bibinfo  {journal} {arXiv preprint arXiv:2112.05332}\ } (\bibinfo {year} {2021})}\BibitemShut {NoStop}%
\bibitem [{\citenamefont {Hotter}\ \emph {et~al.}(2024)\citenamefont {Hotter}, \citenamefont {Ritsch},\ and\ \citenamefont {Gietka}}]{Hotter24}%
  \BibitemOpen
  \bibfield  {author} {\bibinfo {author} {\bibfnamefont {C.}~\bibnamefont {Hotter}}, \bibinfo {author} {\bibfnamefont {H.}~\bibnamefont {Ritsch}},\ and\ \bibinfo {author} {\bibfnamefont {K.}~\bibnamefont {Gietka}},\ }\bibfield  {title} {\bibinfo {title} {Combining critical and quantum metrology},\ }\href {https://doi.org/10.1103/PhysRevLett.132.060801} {\bibfield  {journal} {\bibinfo  {journal} {Phys. Rev. Lett.}\ }\textbf {\bibinfo {volume} {132}},\ \bibinfo {pages} {060801} (\bibinfo {year} {2024})}\BibitemShut {NoStop}%
\bibitem [{\citenamefont {Chen}\ \emph {et~al.}(2024)\citenamefont {Chen}, \citenamefont {L{\"u}}, \citenamefont {Zhu}, \citenamefont {Zhang}, \citenamefont {Ning}, \citenamefont {Yang},\ and\ \citenamefont {Zheng}}]{chen2024critical}%
  \BibitemOpen
  \bibfield  {author} {\bibinfo {author} {\bibfnamefont {K.}~\bibnamefont {Chen}}, \bibinfo {author} {\bibfnamefont {J.-H.}\ \bibnamefont {L{\"u}}}, \bibinfo {author} {\bibfnamefont {X.}~\bibnamefont {Zhu}}, \bibinfo {author} {\bibfnamefont {H.-L.}\ \bibnamefont {Zhang}}, \bibinfo {author} {\bibfnamefont {W.}~\bibnamefont {Ning}}, \bibinfo {author} {\bibfnamefont {Z.-B.}\ \bibnamefont {Yang}},\ and\ \bibinfo {author} {\bibfnamefont {S.-B.}\ \bibnamefont {Zheng}},\ }\bibfield  {title} {\bibinfo {title} {Critical sensing with a single bosonic mode without boson--boson interactions},\ }\href {https://doi.org/10.1002/qute.202400105} {\bibfield  {journal} {\bibinfo  {journal} {Adv. Quantum Technol.}\ }\textbf {\bibinfo {volume} {7}},\ \bibinfo {pages} {2400105} (\bibinfo {year} {2024})}\BibitemShut {NoStop}%
\bibitem [{\citenamefont {Choi}\ \emph {et~al.}(2024)\citenamefont {Choi}, \citenamefont {Salamin}, \citenamefont {Roques-Carmes}, \citenamefont {Sloan}, \citenamefont {Horodynski},\ and\ \citenamefont {Soljacic}}]{choi2024observing}%
  \BibitemOpen
  \bibfield  {author} {\bibinfo {author} {\bibfnamefont {S.}~\bibnamefont {Choi}}, \bibinfo {author} {\bibfnamefont {Y.}~\bibnamefont {Salamin}}, \bibinfo {author} {\bibfnamefont {C.}~\bibnamefont {Roques-Carmes}}, \bibinfo {author} {\bibfnamefont {J.}~\bibnamefont {Sloan}}, \bibinfo {author} {\bibfnamefont {M.}~\bibnamefont {Horodynski}},\ and\ \bibinfo {author} {\bibfnamefont {M.}~\bibnamefont {Soljacic}},\ }\bibfield  {title} {\bibinfo {title} {Observing the dynamics of quantum states generated inside nonlinear optical cavities},\ }\href {https://doi.org/10.48550/arXiv.2412.01772} {\bibfield  {journal} {\bibinfo  {journal} {arXiv preprint arXiv:2412.01772}\ } (\bibinfo {year} {2024})}\BibitemShut {NoStop}%
\bibitem [{\citenamefont {Ilias}\ \emph {et~al.}(2024)\citenamefont {Ilias}, \citenamefont {Yang}, \citenamefont {Huelga},\ and\ \citenamefont {Plenio}}]{Ilias2023}%
  \BibitemOpen
  \bibfield  {author} {\bibinfo {author} {\bibfnamefont {T.}~\bibnamefont {Ilias}}, \bibinfo {author} {\bibfnamefont {D.}~\bibnamefont {Yang}}, \bibinfo {author} {\bibfnamefont {S.~F.}\ \bibnamefont {Huelga}},\ and\ \bibinfo {author} {\bibfnamefont {M.~B.}\ \bibnamefont {Plenio}},\ }\bibfield  {title} {\bibinfo {title} {Criticality-enhanced electric field gradient sensor with single trapped ions},\ }\href {https://doi.org/https://doi.org/10.1038/s41534-024-00833-w} {\bibfield  {journal} {\bibinfo  {journal} {npj Quantum Inf.}\ }\textbf {\bibinfo {volume} {10}},\ \bibinfo {pages} {36} (\bibinfo {year} {2024})}\BibitemShut {NoStop}%
\bibitem [{\citenamefont {Bin}\ \emph {et~al.}(2019)\citenamefont {Bin}, \citenamefont {L\"{u}}, \citenamefont {Yin}, \citenamefont {Zhu}, \citenamefont {Bin},\ and\ \citenamefont {Wu}}]{Bin2019}%
  \BibitemOpen
  \bibfield  {author} {\bibinfo {author} {\bibfnamefont {S.-W.}\ \bibnamefont {Bin}}, \bibinfo {author} {\bibfnamefont {X.-Y.}\ \bibnamefont {L\"{u}}}, \bibinfo {author} {\bibfnamefont {T.-S.}\ \bibnamefont {Yin}}, \bibinfo {author} {\bibfnamefont {G.-L.}\ \bibnamefont {Zhu}}, \bibinfo {author} {\bibfnamefont {Q.}~\bibnamefont {Bin}},\ and\ \bibinfo {author} {\bibfnamefont {Y.}~\bibnamefont {Wu}},\ }\bibfield  {title} {\bibinfo {title} {Mass sensing by quantum criticality},\ }\href {https://doi.org/10.1364/OL.44.000630} {\bibfield  {journal} {\bibinfo  {journal} {Opt. Lett.}\ }\textbf {\bibinfo {volume} {44}},\ \bibinfo {pages} {630} (\bibinfo {year} {2019})}\BibitemShut {NoStop}%
\bibitem [{\citenamefont {Tang}\ \emph {et~al.}(2023)\citenamefont {Tang}, \citenamefont {Qin}, \citenamefont {Liu}, \citenamefont {Wang}, \citenamefont {Cui}, \citenamefont {Su}, \citenamefont {Yan},\ and\ \citenamefont {Chen}}]{Tang2023}%
  \BibitemOpen
  \bibfield  {author} {\bibinfo {author} {\bibfnamefont {S.-B.}\ \bibnamefont {Tang}}, \bibinfo {author} {\bibfnamefont {H.}~\bibnamefont {Qin}}, \bibinfo {author} {\bibfnamefont {B.-B.}\ \bibnamefont {Liu}}, \bibinfo {author} {\bibfnamefont {D.-Y.}\ \bibnamefont {Wang}}, \bibinfo {author} {\bibfnamefont {K.}~\bibnamefont {Cui}}, \bibinfo {author} {\bibfnamefont {S.-L.}\ \bibnamefont {Su}}, \bibinfo {author} {\bibfnamefont {L.-L.}\ \bibnamefont {Yan}},\ and\ \bibinfo {author} {\bibfnamefont {G.}~\bibnamefont {Chen}},\ }\bibfield  {title} {\bibinfo {title} {Enhancement of quantum sensing in a cavity-optomechanical system around the quantum critical point},\ }\href {https://doi.org/10.1103/PhysRevA.108.053514} {\bibfield  {journal} {\bibinfo  {journal} {Phys. Rev. A}\ }\textbf {\bibinfo {volume} {108}},\ \bibinfo {pages} {053514} (\bibinfo {year} {2023})}\BibitemShut {NoStop}%
\bibitem [{\citenamefont {Wan}\ \emph {et~al.}(2024)\citenamefont {Wan}, \citenamefont {Shi},\ and\ \citenamefont {Guan}}]{Wan2023}%
  \BibitemOpen
  \bibfield  {author} {\bibinfo {author} {\bibfnamefont {Q.-K.}\ \bibnamefont {Wan}}, \bibinfo {author} {\bibfnamefont {H.-L.}\ \bibnamefont {Shi}},\ and\ \bibinfo {author} {\bibfnamefont {X.-W.}\ \bibnamefont {Guan}},\ }\bibfield  {title} {\bibinfo {title} {Quantum-enhanced metrology in cavity magnonics},\ }\href {https://doi.org/10.1103/PhysRevB.109.L041301} {\bibfield  {journal} {\bibinfo  {journal} {Phys. Rev. B}\ }\textbf {\bibinfo {volume} {109}},\ \bibinfo {pages} {L041301} (\bibinfo {year} {2024})}\BibitemShut {NoStop}%
\bibitem [{\citenamefont {Mihailescu}\ \emph {et~al.}(2024{\natexlab{a}})\citenamefont {Mihailescu}, \citenamefont {Bayat}, \citenamefont {Campbell},\ and\ \citenamefont {Mitchell}}]{mih23multiparameter}%
  \BibitemOpen
  \bibfield  {author} {\bibinfo {author} {\bibfnamefont {G.}~\bibnamefont {Mihailescu}}, \bibinfo {author} {\bibfnamefont {A.}~\bibnamefont {Bayat}}, \bibinfo {author} {\bibfnamefont {S.}~\bibnamefont {Campbell}},\ and\ \bibinfo {author} {\bibfnamefont {A.~K.}\ \bibnamefont {Mitchell}},\ }\bibfield  {title} {\bibinfo {title} {Multiparameter critical quantum metrology with impurity probes},\ }\href {https://doi.org/10.1088/2058-9565/ad438d} {\bibfield  {journal} {\bibinfo  {journal} {Quantum Sci. Technol.}\ }\textbf {\bibinfo {volume} {9}},\ \bibinfo {pages} {035033} (\bibinfo {year} {2024}{\natexlab{a}})}\BibitemShut {NoStop}%
\bibitem [{\citenamefont {Mihailescu}\ \emph {et~al.}(2024{\natexlab{b}})\citenamefont {Mihailescu}, \citenamefont {Kiely},\ and\ \citenamefont {Mitchell}}]{mihailescu2024}%
  \BibitemOpen
  \bibfield  {author} {\bibinfo {author} {\bibfnamefont {G.}~\bibnamefont {Mihailescu}}, \bibinfo {author} {\bibfnamefont {A.}~\bibnamefont {Kiely}},\ and\ \bibinfo {author} {\bibfnamefont {A.~K.}\ \bibnamefont {Mitchell}},\ }\bibfield  {title} {\bibinfo {title} {Quantum sensing with nanoelectronics: Fisher information for an adiabatic perturbation},\ }\href {https://arxiv.org/abs/2406.18662} {\bibfield  {journal} {\bibinfo  {journal} {arXiv preprint arXiv:2406.18662}\ } (\bibinfo {year} {2024}{\natexlab{b}})}\BibitemShut {NoStop}%
\bibitem [{\citenamefont {Ying}\ \emph {et~al.}(2022)\citenamefont {Ying}, \citenamefont {Felicetti}, \citenamefont {Liu},\ and\ \citenamefont {Braak}}]{Ying2022}%
  \BibitemOpen
  \bibfield  {author} {\bibinfo {author} {\bibfnamefont {Z.-J.}\ \bibnamefont {Ying}}, \bibinfo {author} {\bibfnamefont {S.}~\bibnamefont {Felicetti}}, \bibinfo {author} {\bibfnamefont {G.}~\bibnamefont {Liu}},\ and\ \bibinfo {author} {\bibfnamefont {D.}~\bibnamefont {Braak}},\ }\bibfield  {title} {\bibinfo {title} {Critical quantum metrology in the non-linear quantum {R}abi model},\ }\href {https://doi.org/10.3390/e24081015} {\bibfield  {journal} {\bibinfo  {journal} {Entropy}\ }\textbf {\bibinfo {volume} {24}},\ \bibinfo {pages} {1015} (\bibinfo {year} {2022})}\BibitemShut {NoStop}%
\bibitem [{\citenamefont {Xie}\ \emph {et~al.}(2022)\citenamefont {Xie}, \citenamefont {Xu},\ and\ \citenamefont {Wang}}]{Xie2022}%
  \BibitemOpen
  \bibfield  {author} {\bibinfo {author} {\bibfnamefont {D.}~\bibnamefont {Xie}}, \bibinfo {author} {\bibfnamefont {C.}~\bibnamefont {Xu}},\ and\ \bibinfo {author} {\bibfnamefont {A.~M.}\ \bibnamefont {Wang}},\ }\bibfield  {title} {\bibinfo {title} {Quantum thermometry with a dissipative quantum {R}abi system},\ }\href {https://doi.org/10.1140/epjp/s13360-022-03524-7} {\bibfield  {journal} {\bibinfo  {journal} {Eur. Phys. J. Plus}\ }\textbf {\bibinfo {volume} {137}},\ \bibinfo {pages} {1323} (\bibinfo {year} {2022})}\BibitemShut {NoStop}%
\bibitem [{\citenamefont {L\"u}\ \emph {et~al.}(2022)\citenamefont {L\"u}, \citenamefont {Ning}, \citenamefont {Zhu}, \citenamefont {Wu}, \citenamefont {Shen}, \citenamefont {Yang},\ and\ \citenamefont {Zheng}}]{Lu2022}%
  \BibitemOpen
  \bibfield  {author} {\bibinfo {author} {\bibfnamefont {J.-H.}\ \bibnamefont {L\"u}}, \bibinfo {author} {\bibfnamefont {W.}~\bibnamefont {Ning}}, \bibinfo {author} {\bibfnamefont {X.}~\bibnamefont {Zhu}}, \bibinfo {author} {\bibfnamefont {F.}~\bibnamefont {Wu}}, \bibinfo {author} {\bibfnamefont {L.-T.}\ \bibnamefont {Shen}}, \bibinfo {author} {\bibfnamefont {Z.-B.}\ \bibnamefont {Yang}},\ and\ \bibinfo {author} {\bibfnamefont {S.-B.}\ \bibnamefont {Zheng}},\ }\bibfield  {title} {\bibinfo {title} {Critical quantum sensing based on the {J}aynes-{C}ummings model with a squeezing drive},\ }\href {https://doi.org/10.1103/PhysRevA.106.062616} {\bibfield  {journal} {\bibinfo  {journal} {Phys. Rev. A}\ }\textbf {\bibinfo {volume} {106}},\ \bibinfo {pages} {062616} (\bibinfo {year} {2022})}\BibitemShut {NoStop}%
\bibitem [{\citenamefont {Zhu}\ \emph {et~al.}(2024)\citenamefont {Zhu}, \citenamefont {L\"u}, \citenamefont {Ning}, \citenamefont {Shen}, \citenamefont {Wu},\ and\ \citenamefont {Yang}}]{Zhu24}%
  \BibitemOpen
  \bibfield  {author} {\bibinfo {author} {\bibfnamefont {X.}~\bibnamefont {Zhu}}, \bibinfo {author} {\bibfnamefont {J.-H.}\ \bibnamefont {L\"u}}, \bibinfo {author} {\bibfnamefont {W.}~\bibnamefont {Ning}}, \bibinfo {author} {\bibfnamefont {L.-T.}\ \bibnamefont {Shen}}, \bibinfo {author} {\bibfnamefont {F.}~\bibnamefont {Wu}},\ and\ \bibinfo {author} {\bibfnamefont {Z.-B.}\ \bibnamefont {Yang}},\ }\bibfield  {title} {\bibinfo {title} {Quantum geometric tensor and critical metrology in the anisotropic {D}icke model},\ }\href {https://doi.org/10.1103/PhysRevA.109.052621} {\bibfield  {journal} {\bibinfo  {journal} {Phys. Rev. A}\ }\textbf {\bibinfo {volume} {109}},\ \bibinfo {pages} {052621} (\bibinfo {year} {2024})}\BibitemShut {NoStop}%
\bibitem [{\citenamefont {Demkowicz-Dobrza\ifmmode~\acute{n}\else \'{n}\fi{}ski}\ and\ \citenamefont {Maccone}(2014)}]{Maccone_entangl}%
  \BibitemOpen
  \bibfield  {author} {\bibinfo {author} {\bibfnamefont {R.}~\bibnamefont {Demkowicz-Dobrza\ifmmode~\acute{n}\else \'{n}\fi{}ski}}\ and\ \bibinfo {author} {\bibfnamefont {L.}~\bibnamefont {Maccone}},\ }\bibfield  {title} {\bibinfo {title} {Using entanglement against noise in quantum metrology},\ }\href {https://doi.org/10.1103/PhysRevLett.113.250801} {\bibfield  {journal} {\bibinfo  {journal} {Phys. Rev. Lett.}\ }\textbf {\bibinfo {volume} {113}},\ \bibinfo {pages} {250801} (\bibinfo {year} {2014})}\BibitemShut {NoStop}%
\bibitem [{\citenamefont {Kurdzia\l{}ek}\ \emph {et~al.}(2023)\citenamefont {Kurdzia\l{}ek}, \citenamefont {G\'orecki}, \citenamefont {Albarelli},\ and\ \citenamefont {Demkowicz-Dobrza\ifmmode~\acute{n}\else \'{n}\fi{}ski}}]{Gorecki_bounds}%
  \BibitemOpen
  \bibfield  {author} {\bibinfo {author} {\bibfnamefont {S.}~\bibnamefont {Kurdzia\l{}ek}}, \bibinfo {author} {\bibfnamefont {W.}~\bibnamefont {G\'orecki}}, \bibinfo {author} {\bibfnamefont {F.}~\bibnamefont {Albarelli}},\ and\ \bibinfo {author} {\bibfnamefont {R.}~\bibnamefont {Demkowicz-Dobrza\ifmmode~\acute{n}\else \'{n}\fi{}ski}},\ }\bibfield  {title} {\bibinfo {title} {Using adaptiveness and causal superpositions against noise in quantum metrology},\ }\href {https://doi.org/10.1103/PhysRevLett.131.090801} {\bibfield  {journal} {\bibinfo  {journal} {Phys. Rev. Lett.}\ }\textbf {\bibinfo {volume} {131}},\ \bibinfo {pages} {090801} (\bibinfo {year} {2023})}\BibitemShut {NoStop}%
\bibitem [{\citenamefont {Demkowicz-Dobrza\ifmmode~\acute{n}\else \'{n}\fi{}ski}\ \emph {et~al.}(2017)\citenamefont {Demkowicz-Dobrza\ifmmode~\acute{n}\else \'{n}\fi{}ski}, \citenamefont {Czajkowski},\ and\ \citenamefont {Sekatski}}]{demkowicz2017adaptive}%
  \BibitemOpen
  \bibfield  {author} {\bibinfo {author} {\bibfnamefont {R.}~\bibnamefont {Demkowicz-Dobrza\ifmmode~\acute{n}\else \'{n}\fi{}ski}}, \bibinfo {author} {\bibfnamefont {J.}~\bibnamefont {Czajkowski}},\ and\ \bibinfo {author} {\bibfnamefont {P.}~\bibnamefont {Sekatski}},\ }\bibfield  {title} {\bibinfo {title} {Adaptive quantum metrology under general {M}arkovian noise},\ }\href {https://doi.org/10.1103/PhysRevX.7.041009} {\bibfield  {journal} {\bibinfo  {journal} {Phys. Rev. X}\ }\textbf {\bibinfo {volume} {7}},\ \bibinfo {pages} {041009} (\bibinfo {year} {2017})}\BibitemShut {NoStop}%
\bibitem [{\citenamefont {G{\'o}recki}\ \emph {et~al.}(2024)\citenamefont {G{\'o}recki}, \citenamefont {Albarelli}, \citenamefont {Felicetti}, \citenamefont {Di~Candia},\ and\ \citenamefont {Maccone}}]{Gorecki24}%
  \BibitemOpen
  \bibfield  {author} {\bibinfo {author} {\bibfnamefont {W.}~\bibnamefont {G{\'o}recki}}, \bibinfo {author} {\bibfnamefont {F.}~\bibnamefont {Albarelli}}, \bibinfo {author} {\bibfnamefont {S.}~\bibnamefont {Felicetti}}, \bibinfo {author} {\bibfnamefont {R.}~\bibnamefont {Di~Candia}},\ and\ \bibinfo {author} {\bibfnamefont {L.}~\bibnamefont {Maccone}},\ }\bibfield  {title} {\bibinfo {title} {Interplay between time and energy in bosonic noisy quantum metrology},\ }\href {https://doi.org/10.48550/arXiv.2409.18791} {\bibfield  {journal} {\bibinfo  {journal} {arXiv preprint arXiv:2409.18791}\ } (\bibinfo {year} {2024})}\BibitemShut {NoStop}%
\bibitem [{\citenamefont {Calvanese~Strinati}\ and\ \citenamefont {Conti}(2024)}]{Calvanese24}%
  \BibitemOpen
  \bibfield  {author} {\bibinfo {author} {\bibfnamefont {M.}~\bibnamefont {Calvanese~Strinati}}\ and\ \bibinfo {author} {\bibfnamefont {C.}~\bibnamefont {Conti}},\ }\bibfield  {title} {\bibinfo {title} {Non-{G}aussianity in the quantum parametric oscillator},\ }\href {https://doi.org/10.1103/PhysRevA.109.063519} {\bibfield  {journal} {\bibinfo  {journal} {Phys. Rev. A}\ }\textbf {\bibinfo {volume} {109}},\ \bibinfo {pages} {063519} (\bibinfo {year} {2024})}\BibitemShut {NoStop}%
\bibitem [{\citenamefont {Mihailescu}\ \emph {et~al.}(2024{\natexlab{c}})\citenamefont {Mihailescu}, \citenamefont {Campbell},\ and\ \citenamefont {Gietka}}]{Mihailescu24}%
  \BibitemOpen
  \bibfield  {author} {\bibinfo {author} {\bibfnamefont {G.}~\bibnamefont {Mihailescu}}, \bibinfo {author} {\bibfnamefont {S.}~\bibnamefont {Campbell}},\ and\ \bibinfo {author} {\bibfnamefont {K.}~\bibnamefont {Gietka}},\ }\bibfield  {title} {\bibinfo {title} {Uncertain quantum critical metrology: From single to multi parameter sensing},\ }\href {https://arxiv.org/abs/2407.19917} {\bibfield  {journal} {\bibinfo  {journal} {arXiv preprint arXiv:2407.19917}\ } (\bibinfo {year} {2024}{\natexlab{c}})}\BibitemShut {NoStop}%
\bibitem [{\citenamefont {Carusotto}\ and\ \citenamefont {Ciuti}(2013)}]{RevModPhys.85.299}%
  \BibitemOpen
  \bibfield  {author} {\bibinfo {author} {\bibfnamefont {I.}~\bibnamefont {Carusotto}}\ and\ \bibinfo {author} {\bibfnamefont {C.}~\bibnamefont {Ciuti}},\ }\bibfield  {title} {\bibinfo {title} {Quantum fluids of light},\ }\href {https://doi.org/10.1103/RevModPhys.85.299} {\bibfield  {journal} {\bibinfo  {journal} {Rev. Mod. Phys.}\ }\textbf {\bibinfo {volume} {85}},\ \bibinfo {pages} {299} (\bibinfo {year} {2013})}\BibitemShut {NoStop}%
\bibitem [{\citenamefont {Barzanjeh}\ \emph {et~al.}(2022)\citenamefont {Barzanjeh}, \citenamefont {Xuereb}, \citenamefont {Gr{\"o}blacher}, \citenamefont {Paternostro}, \citenamefont {Regal},\ and\ \citenamefont {Weig}}]{barzanjeh2022optomechanics}%
  \BibitemOpen
  \bibfield  {author} {\bibinfo {author} {\bibfnamefont {S.}~\bibnamefont {Barzanjeh}}, \bibinfo {author} {\bibfnamefont {A.}~\bibnamefont {Xuereb}}, \bibinfo {author} {\bibfnamefont {S.}~\bibnamefont {Gr{\"o}blacher}}, \bibinfo {author} {\bibfnamefont {M.}~\bibnamefont {Paternostro}}, \bibinfo {author} {\bibfnamefont {C.~A.}\ \bibnamefont {Regal}},\ and\ \bibinfo {author} {\bibfnamefont {E.~M.}\ \bibnamefont {Weig}},\ }\bibfield  {title} {\bibinfo {title} {Optomechanics for quantum technologies},\ }\href {https://www.nature.com/articles/s41567-021-01402-0} {\bibfield  {journal} {\bibinfo  {journal} {Nat. Phys.}\ }\textbf {\bibinfo {volume} {18}},\ \bibinfo {pages} {15} (\bibinfo {year} {2022})}\BibitemShut {NoStop}%
\bibitem [{\citenamefont {Slim}\ \emph {et~al.}(2024)\citenamefont {Slim}, \citenamefont {Wanjura}, \citenamefont {Brunelli}, \citenamefont {del Pino}, \citenamefont {Nunnenkamp},\ and\ \citenamefont {Verhagen}}]{Slim_2024}%
  \BibitemOpen
  \bibfield  {author} {\bibinfo {author} {\bibfnamefont {J.~J.}\ \bibnamefont {Slim}}, \bibinfo {author} {\bibfnamefont {C.~C.}\ \bibnamefont {Wanjura}}, \bibinfo {author} {\bibfnamefont {M.}~\bibnamefont {Brunelli}}, \bibinfo {author} {\bibfnamefont {J.}~\bibnamefont {del Pino}}, \bibinfo {author} {\bibfnamefont {A.}~\bibnamefont {Nunnenkamp}},\ and\ \bibinfo {author} {\bibfnamefont {E.}~\bibnamefont {Verhagen}},\ }\bibfield  {title} {\bibinfo {title} {Optomechanical realization of the bosonic {K}itaev chain},\ }\href {https://doi.org/10.1038/s41586-024-07174-w} {\bibfield  {journal} {\bibinfo  {journal} {Nature}\ }\textbf {\bibinfo {volume} {627}},\ \bibinfo {pages} {767–771} (\bibinfo {year} {2024})}\BibitemShut {NoStop}%
\bibitem [{\citenamefont {Gu}\ \emph {et~al.}(2017)\citenamefont {Gu}, \citenamefont {Kockum}, \citenamefont {Miranowicz}, \citenamefont {xi~Liu},\ and\ \citenamefont {Nori}}]{GU20171}%
  \BibitemOpen
  \bibfield  {author} {\bibinfo {author} {\bibfnamefont {X.}~\bibnamefont {Gu}}, \bibinfo {author} {\bibfnamefont {A.~F.}\ \bibnamefont {Kockum}}, \bibinfo {author} {\bibfnamefont {A.}~\bibnamefont {Miranowicz}}, \bibinfo {author} {\bibfnamefont {Y.}~\bibnamefont {xi~Liu}},\ and\ \bibinfo {author} {\bibfnamefont {F.}~\bibnamefont {Nori}},\ }\bibfield  {title} {\bibinfo {title} {Microwave photonics with superconducting quantum circuits},\ }\href {https://doi.org/https://doi.org/10.1016/j.physrep.2017.10.002} {\bibfield  {journal} {\bibinfo  {journal} {Phys. Rep.}\ }\textbf {\bibinfo {volume} {718-719}},\ \bibinfo {pages} {1} (\bibinfo {year} {2017})}\BibitemShut {NoStop}%
\bibitem [{\citenamefont {McDonald}\ and\ \citenamefont {Clerk}(2020)}]{mcdonald2020exponentially}%
  \BibitemOpen
  \bibfield  {author} {\bibinfo {author} {\bibfnamefont {A.}~\bibnamefont {McDonald}}\ and\ \bibinfo {author} {\bibfnamefont {A.~A.}\ \bibnamefont {Clerk}},\ }\bibfield  {title} {\bibinfo {title} {Exponentially-enhanced quantum sensing with non-{H}ermitian lattice dynamics},\ }\href {https://doi.org/https://doi.org/10.1038/s41467-020-19090-4} {\bibfield  {journal} {\bibinfo  {journal} {Nat. Commun.}\ }\textbf {\bibinfo {volume} {11}},\ \bibinfo {pages} {5382} (\bibinfo {year} {2020})}\BibitemShut {NoStop}%
\bibitem [{\citenamefont {Grabner}\ and\ \citenamefont {Prodinger}(2007)}]{grabner2007secant}%
  \BibitemOpen
  \bibfield  {author} {\bibinfo {author} {\bibfnamefont {P.~J.}\ \bibnamefont {Grabner}}\ and\ \bibinfo {author} {\bibfnamefont {H.}~\bibnamefont {Prodinger}},\ }\bibfield  {title} {\bibinfo {title} {Secant and cosecant sums and {B}ernoulli-{N}{\"o}rlund polynomials},\ }\href {https://doi.org/10.2989/16073600709486191} {\bibfield  {journal} {\bibinfo  {journal} {Quaest. Math.}\ }\textbf {\bibinfo {volume} {30}},\ \bibinfo {pages} {159} (\bibinfo {year} {2007})}\BibitemShut {NoStop}%
\bibitem [{\citenamefont {Blagouchine}\ and\ \citenamefont {Moreau}(2024)}]{blagouchine2024finite}%
  \BibitemOpen
  \bibfield  {author} {\bibinfo {author} {\bibfnamefont {I.~V.}\ \bibnamefont {Blagouchine}}\ and\ \bibinfo {author} {\bibfnamefont {E.}~\bibnamefont {Moreau}},\ }\bibfield  {title} {\bibinfo {title} {On a finite sum of cosecants appearing in various problems},\ }\href {https://doi.org/10.1016/j.jmaa.2024.128515} {\bibfield  {journal} {\bibinfo  {journal} {J. Math. Anal. Appl.}\ }\textbf {\bibinfo {volume} {539}},\ \bibinfo {pages} {128515} (\bibinfo {year} {2024})}\BibitemShut {NoStop}%
\end{thebibliography}%

\end{document}